%% file: sample-sigconf.tex
\documentclass[sigconf]{acmart}

\usepackage[ruled]{algorithm2e}
\usepackage{multirow}

\usepackage{enumitem}
\AtBeginDocument{%
  \providecommand\BibTeX{{%
    \normalfont B\kern-0.5em{\scshape i\kern-0.25em b}\kern-0.8em\TeX}}}

\setcopyright{acmcopyright}
\copyrightyear{2021}
\acmYear{2021}
\acmDOI{10.1145/1122445.1122456}

\copyrightyear{2021}
\acmYear{2021}
\setcopyright{acmlicensed}\acmConference[CIKM '21]{Proceedings of the 30th ACM International Conference on Information and Knowledge Management}{November 1--5, 2021}{Virtual Event, QLD, Australia}
\acmBooktitle{Proceedings of the 30th ACM International Conference on Information and Knowledge Management (CIKM '21), November 1--5, 2021, Virtual Event, QLD, Australia}
\acmPrice{15.00}
\acmDOI{10.1145/3459637.3482463}
\acmISBN{978-1-4503-8446-9/21/11}

\begin{document}

\fancyhead{}

\title{POSSCORE: A Simple Yet Effective Evaluation of \\Conversational Search with Part of Speech Labelling}

\author{Zeyang Liu}
\email{zeyang.liu@nottingham.ac.uk}
\affiliation{%
  \institution{University of Nottingham}
  \streetaddress{Jubilee Campus, Wollaton Road}
  \city{Nottingham}
  \country{UK}
  \postcode{NG8 1BB}
}

\author{Ke Zhou}
\email{ke.zhou@nottingham.ac.uk}
\affiliation{%
  \institution{University of Nottingham \& Nokia Bell Labs}
  \streetaddress{Jubilee Campus, Wollaton Road}
  \city{Nottingham}
  \country{UK}
  \postcode{NG8 1BB}
}

\author{Jiaxin Mao}
\email{maojiaxin@ruc.edu.cn}
\affiliation{%
  \institution{Renmin University of China}
  \streetaddress{}
  \city{Beijing}
  \country{China}
  \postcode{100872}
}

\author{Max L. Wilson}
\email{max.wilson@nottingham.ac.uk}
\orcid{0000-0002-3515-6633}
\affiliation{%
  \institution{University of Nottingham}
  \streetaddress{Jubilee Campus, Wollaton Road}
  \city{Nottingham}
  \country{UK}
  \postcode{NG8 1BB}
}

\renewcommand{\shortauthors}{Liu et al.}

\begin{abstract}
  Conversational search systems, such as Google Assistant and Microsoft Cortana, provide a new search paradigm where users are allowed, via natural language dialogues, to communicate with search systems. Evaluating such systems is very challenging since search results are presented in the format of natural language sentences. Given the unlimited number of possible responses, collecting relevance assessments for all the possible responses is infeasible. In this paper, we propose POSSCORE \footnote{The codes are available at \url{https://github.com/zy-liu/POSSCORE} }, a simple yet effective automatic evaluation method for conversational search. The proposed embedding-based metric takes the influence of part of speech (POS) of the terms in the response into account. To the best knowledge, our work is the first to systematically demonstrate the importance of incorporating \emph{syntactic} information, such as POS labels, for conversational search evaluation. Experimental results demonstrate that our metrics can correlate with human preference, achieving significant improvements over state-of-the-art baseline metrics. 
\end{abstract}





\maketitle

\section{Introduction} \label{sec:intro}
\input{content/1_introduction.tex}

\section{Related Work} \label{sec:related_work}
\input{content/2_related_work.tex}
\section{Proposed Metrics} \label{sec:proposed_metric}
\input{content/3_0_proposed_metrics.tex}
\section{Experimental Settings} \label{sec:experiment_settings}
\input{content/4_0_experimental_settings.tex}
\section{Experimental Results} \label{sec:results}
\input{content/5_0_experiment_results.tex}

%\section{Discussions} \label{sec:discussion}
%\input{content/6_discussion.tex}
\section{Conclusions}  \label{sec:conclusion}
\input{content/6_conclusion.tex}
\bibliographystyle{ACM-Reference-Format}
\bibliography{sample-base}

\appendix

\end{document}

%% file: content/1_introduction.tex
Various existing voice assistants and automatic desk helpers, such as Apple Siri, Google Assistant, and Microsoft Cortana, can be seen as typical platforms for conversational search. 
The major difference between conversational search and traditional search is that conversational search systems expect users to describe their search tasks using natural language and typically presents the search results through conversations. However, it should be noted that the core of conversational search is still an information retrieval~(IR) system. In other words, the goal of conversational search is to seek information and satisfy users' information needs \cite{radlinski2017theoretical, vtyurina2017exploring}.

\begin{table*}[t] \footnotesize
\caption{An example from the PersonaChat dataset \cite{zhang2018personalizing}. The selected POS tags of our proposed POSSCORE are <`ADJ', `ADV', `VERB', `PROPN', `NOUN'>. It can be observed that our POSSCORE metric aligns better with human perceived quality score than BLEU4 and METEOR. Higher metric score means the corresponding metric is more likely to prefer this candidate.}
\vspace{-0.1in}
\begin{tabular}{c|lcccc}
\hline
\multicolumn{1}{l|}{} & \multicolumn{5}{l}{...} \\
Context & \multicolumn{5}{l}{A: My dad had me hooked on high protein foods like fish and chicken when I was training.} \\
 & \multicolumn{5}{l}{B: What are you training for?} \\ \hline
Reference & \multicolumn{5}{l}{I am competing {\color{blue}(VERB)} for a national {\color{blue}(ADJ)} chess {\color{blue}(NOUN)} tournament {\color{blue}(NOUN)}. It helps {\color{blue}(VERB)} me keep {\color{blue}(VERB)} focus {\color{blue}(NOUN)}.} \\ \hline
Good candidate & \multicolumn{5}{p{13cm}}{Chess {\color{blue}(NOUN)} tournaments{\color{blue}(NOUN)} protein {\color{blue}(NOUN)} makes {\color{blue}(VERB)} your brain {\color{blue}{(NOUN)}} more {\color{blue}{(ADV)}} activate {\color{blue}(VERB)} for intense {\color{blue}{(ADJ)}} chess {\color{blue}{(NOUN)}} matches {\color{blue}{(NOUN)}.}} \\ \hline
Bad candidate & \multicolumn{5}{l}{I am a professional {\color{blue}(ADJ)} chess {\color{blue}(NOUN)} player {\color{blue}(NOUN)}. }  \\ \hline
  & \multicolumn{1}{c|}{BLEU4} &  \multicolumn{1}{c|}{METEOR} &  \multicolumn{1}{c|}{BERT-Score} & \multicolumn{1}{c|}{POSSCORE} & Human score (5-point scale: 0 - poor quality, 5 - excellent quality) \\ \hline
Good candidate & \multicolumn{1}{c|}{0.019} & \multicolumn{1}{c|}{0.155} & \multicolumn{1}{c|}{0.859} & \multicolumn{1}{c|}{\textbf{1.942}} & \textbf{5}\\
Bad candidate & \multicolumn{1}{c|}{\textbf{0.032}} & \multicolumn{1}{c|}{\textbf{0.257}} & \multicolumn{1}{c|}{\textbf{0.892}}& \multicolumn{1}{c|}{1.476} & 2\\ \hline
\end{tabular}

\label{tab:intro_example}
\vspace{-0.1in}
\end{table*}

Evaluation plays a pivotal role in designing and tuning search systems \cite{clarke2008novelty}. However, due to the nature of the conversational search,
%due to the rapid evolution of interaction mode in conversational search, 
it is difficult to apply traditional evaluation methods, based on relevance assessments, to this new search paradigm.
%in this new search scenairo
%traditional search evaluation methods, such as TF-IDF 
%\mjx{(I don't think TF-IDF is an evaluation method, maybe we could say ``it might be difficult to apply the traditional evaluation methods based on relevance assessments in this new search scenairo'')}
%, might be difficult to measure the relevance of the search results. 
%Since users directly interact with systems within utterance, the lack of explicit and implicit feedback is the major challenge of the evaluation process. \mjx{I think here we should be more specific about why existing evaluation does not work here: ``When users directly interact with conversational search systems in natural language, the numbers of possible user utterances and system responses are infinite, which makes it hard to collect a reusable set of assessments for the evaluation of conversational search. ''} 
When users directly interact with conversational search systems in natural language, the numbers of possible user utterances and system responses are infinite, which makes it hard to collect a set of reusable assessments for the evaluation of conversational search. To face such challenges, previous work has proposed a number of automatic evaluation metrics to quantify the semantic similarity of utterances to references (ground-truth responses) and leverage this as the proxy for relevance. Examples of those metrics include word-overlap based measures (e.g.,~BLEU \cite{papineni2002bleu}, METEOR \cite{banerjee2005meteor}) , word-embedding based metrics (e.g.,~Embedding Average \cite{serban2017hierarchical}, Soft Cosine Similarity \cite{sidorov2014soft} and BERTScore \cite{bert-score}), and learning-based metrics (e.g., BERT-RUBER \cite{ghazarian2019better}). However, %except for BERTScore, 
most of the above metrics use the entire sentence as the input and treat all the words of the responses equally in the evaluation process, which inevitably brings much noise in estimating relevance. A few prior meta-evaluation studies \cite{liu2016not, novikova2017we, liu2021meta} have revealed the weaknesses of existing automatic metrics.
Further, empirical studies have demonstrated that all of these metrics correlate weakly with human preference.
%both the word-overlap based (e.g.,~BLEU) and word-embedding based (e.g.,~Embedding Average) metrics correlate weakly with human judgements in conversational search systems.

The \emph{syntactic structures} of the utterances might capture additional information for evaluation.
The part of speech (POS) defines how a word is used in a sentence and what role the word plays within the grammatical structure of phrases. Usually, words with the same POS tags contain related grammatical functions and display similar semantic behaviour. \citet{novikova2017we} have demonstrated that grammar-based metrics correlate better with quality for evaluation in the area of natural language generation. 
POS words and labels, which contain grammatical information, might be helpful for utterance evaluation in conversational search. 
Therefore, in this paper, we comprehensively analyze the effect of POS words and labels in the evaluation process.
To our best knowledge, this is the first work to systematically demonstrate the importance of the part of speech labels for conversational search evaluation.
An evaluation example can be found in Table ~\ref{tab:intro_example}, which demonstrates the effectiveness of our proposed metric. Compared to BLEU metrics that utilize only semantic matching, our proposed metric exploits the syntactic POS distributions. 
Given that the POS distributions of the good candidate result align better with that of the reference (ground truth) than a bad candidate, our proposed metric POSSCORE effectively captures such syntactic matching and correlates better with the human perceived quality score.
%\todo{present an example demonstrating POS based approach is better than traditional BLEU for example}
%Please see Table ~\ref{tab:intro_example}
%\todo{focus on single-turn conversational search}
%The main contributions of our work are two-fold:

Our main contributions are two-fold:
%\begin{itemize}
    %\item 
    (1) We are the first to systematically reveal the connection between POS labels and relevance in conversational search evaluation. We empirically demonstrate that conversational search evaluation should also consider syntactic information, such as POS, rather than only the words in responses.
    %\item 
    (2) We propose a simple yet effective POS-based metric: POSSCORE. Experimental results show our metrics can correlate strongly with human preference, outperforming state-of-the-art baselines.
%\end{itemize}

\begin{comment}
The paper is organized as follows.
\S\ref{sec:related_work} presents previous work, describing existing evaluation metrics. Our proposed metrics are presented in \S\ref{sec:proposed_metric}. The experiment settings are described in \S\ref{sec:experiment_settings}.  \S\ref{sec:results} shows the experiment results of our proposed metrics. A discussion and conclusion of this work are described in \S\ref{sec:discussion} and \S\ref{sec:conclusion}.
\end{comment}

%% file: content/2_related_work.tex
%\subsection{Evaluation of Conversational Search}
%\label{sec:evaluation_of_conversational_search}

\noindent {\bf Evaluation of Conversational Search} With the lack of a uniform structure, as created by traditional search systems, it is challenging to find suitable features to capture the quality of responses in conversational search evaluation \cite{lipani2021doing}. To address this problem, the general idea of prior evaluation studies is to evaluate the appropriateness of responses by comparing the candidate system responses with ideal ones, which are usually generated by humans. With the similarity between ideal responses (i.e., ground truth or reference response) and candidate responses, automatic metrics can estimate a quality score of candidate responses. As far as we know, there are generally three categories of metrics in conversational search evaluation: word overlap-based metrics, word embedding-based metrics, and learning-based metrics.

%\begin{itemize}
    %\item 
    \noindent \textit{Word overlap-based metrics.} The basic idea of these metrics is to compute the number of overlapping words between references and candidate responses. Since this type of metric usually has simple algorithms and interpretable structures, word overlap-based metrics have become a popular choice for conversational search and open-domain dialogue evaluation. Typical metrics such as BLEU \cite{papineni2002bleu} and METEOR \cite{banerjee2005meteor} have been widely used to evaluate the adequacy of a response, especially in dialogue-related competitions\cite{kim2019eighth}. However, previous studies \cite{liu2016not,novikova2017we} indicate that these overlap-based metrics weakly correlate with human judgements.
    
    %\item 
    \noindent \textit{Word embedding-based metrics.} The shortfall of word overlap-based metrics is obvious: the exact matching methods are not able to capture the potential connection between words that are similar topically. Therefore, embedding-based metrics are proposed to address this issue. Popular metrics such as Greedy Matching \cite{rus2012comparison}, Vector Extrema\cite{forgues2014bootstrapping} and BERTScore \cite{bert-score} are also widely applied in dialogue evaluation \cite{lan2019talk, mehri2020usr}.
    
    %\item 
    \noindent \textit{Learning-based metrics.} The basic idea of these metrics is to train a supervised model to learn the underlying criteria of human judgements. The features adopted in training models can be the semantic features of ground truth or the context. For example, \citet{lowe2017towards} propose ADEM, which is a recurrent neural network model, to fit the ratings of human judgements. \citet{tao2018ruber} proposed a mixed evaluation method combining referenced and unreferenced metrics. Especially, their unreferenced part is a supervised model, which aims to estimate the appropriateness of response with respect to the context. Although this kind of metric can achieve good performance in some specific scenarios, their training process is inevitably influenced by the given training datasets. In other words, the evaluation score may be different, even when we test the same dataset, if we use different training settings. Further, it is difficult to interpret the results of learning-based metrics. 
    
%\end{itemize}

Besides the above ``offline'' evaluation methods, online methods, such as satisfaction prediction, are also popular in conversational search evaluation. Different from offline methods, the online evaluation focuses on users' behaviour and feedback when interacting with systems in real-time. Many prior studies have presented methods for satisfaction prediction for intelligent assistants, such as \cite{kiseleva2016predicting, kiseleva2016understanding, hashemi2018measuring}. The basic idea of these methods is to construct a predictive model based on user interaction behaviour signals or semantic features, and estimate a score of an utterance that is close to human judgements.
In this paper, we aim to propose a simple POS-based evaluation metric for conversational search. Since the corpora we used do not contain adequate user interaction information, our work only considers offline evaluation methods as baselines.

%\subsection{Predictive Power}

\noindent {\bf Predictive Power} To examine the efficiency and fidelity of a metric, one common approach is to compare the correlation rate between human annotation scores and the metric scores  \cite{bert-score, lan2020pone, liu2016not}. However, the correlation score might be invalid when there are many ties (i.e., the values are the same in each pair) in the datasets. In our paper, we choose to adopt \emph{predictive power}, which measures the agreement between user preferences and metrics when presenting a pair of different responses, to evaluate the fidelity of proposed metrics. Many prior studies have demonstrated its suitability for measuring the fidelity of metrics in meta-evaluation \cite{sanderson2010user, zhou2012evaluating, sakai2012evaluation, chen2017meta, sakai2019diversity, sakai2005effect, liu2021meta}.

%% file: content/3_0_proposed_metrics.tex
This section focuses on the methodology of our proposed metrics. First, the methods and analysis for Part of Speech (POS) labelling are presented in \S\ref{sec:POS_labelling} and \S\ref{sec:POS_analysis}. The adaptive extension based on existing metrics, such as BLEU and METEOR, are introduced in \S\ref{sec:incorporation_method}. We describe the design of our POSSCORE metric in \S\ref{sec:posscore_metric}.

\subsection{Part of Speech Labelling}
\label{sec:POS_labelling}
\input{content/3_1_part_of_speech_labelling.tex}

\subsection{Analysis of POS Words in the Responses}
\label{sec:POS_analysis}
\input{content/3_2_analysis_of_POS_word.tex}

\subsection{POS-aware Adapted Metrics} \label{sec:incorporation_method}
\input{content/3_3_incorporation_method}

\subsection{Proposed POSSCORE Metric}
\label{sec:posscore_metric}
\input{content/3_4_posscore_metric}

%% file: content/3_1_part_of_speech_labelling.tex
The part of speech (POS) explains how a word is used in a sentence and what role it plays within the grammatical structure of the sentence. Words with the same POS tags typically contain identical grammatical information and exhibit similar semantic behaviour. Therefore, using appropriate POS tags allows us to bring grammatical information into the automatic evaluation and to potentially achieve better evaluation performance.
In the field of natural language processing, many previous studies have proposed efficient and effective POS tagging methods, including rule-based methods (e.g., \cite{brill1992simple}) and learning-based methods (e.g., \cite{ratnaparkhi1996maximum, kupiec1992robust}). In our study, we adopt the spaCy toolkit\footnote{https://github.com/explosion/spaCy}, a popular industrial NLP library, to extract the POS tags from the responses. Table \ref{tab:pos_labels} shows the categories of POS tags in the spaCy toolkit and which tags were used in our experiment. Here, we select POS tags according to two criteria: (1) \emph{informativeness}: POS tags should have factual information; (2) \emph{interpretability}: selected POS tags can be directly interpreted, which ensures our designed metrics could be interpretable and extensible. Thus, we finally choose ADJ, ADV, VERB, NOUN, PRON, and PROPN as the candidate POS tags.  We further refer to the words with these informative and interpretable POS tags as \emph{POS words}. 

%\todo{(minor) change table 3, adding example words for a given category, put ``not-adopted'' ones at the bottom part of the table }

\begin{table}[] 
\footnotesize
\caption{The POS tags in spaCy. `Adopted POS tags' shows the POS tags used in our experiment and `NOT adopted POS tags' presents the rest POS tags in spaCy.}
\vspace{-0.1in}
\begin{tabular}{cc|cc}
\hline
\multicolumn{2}{c|}{Adopted POS tags} & \multicolumn{2}{c}{NOT adopted POS tags} \\ \hline
POS tag & Description & POS tag & Description \\ \hline
ADJ & adjective & AUX & auxiliary verb \\
ADV & adverb & CONJ & coordinating conjunction \\
VERB & verb & DET & determiner \\
NOUN & noun & INTJ & interjection \\
PRON & pronoun & NUM & numeral \\
PROPN & proper noun & PART & particle \\
 &  & PUNCT & punctuation \\
 &  & SCONJ & subordinating conjunction \\
 &  & SYM & symbol \\
 &  & ADP & adposition \\
 &  & X & other \\ \hline
\end{tabular}
\label{tab:pos_labels}
\vspace{-0.1in}
\end{table}

%% file: content/3_2_analysis_of_POS_word.tex
\begin{figure*}[]
    \footnotesize
    \centering
    \includegraphics[width=\textwidth]{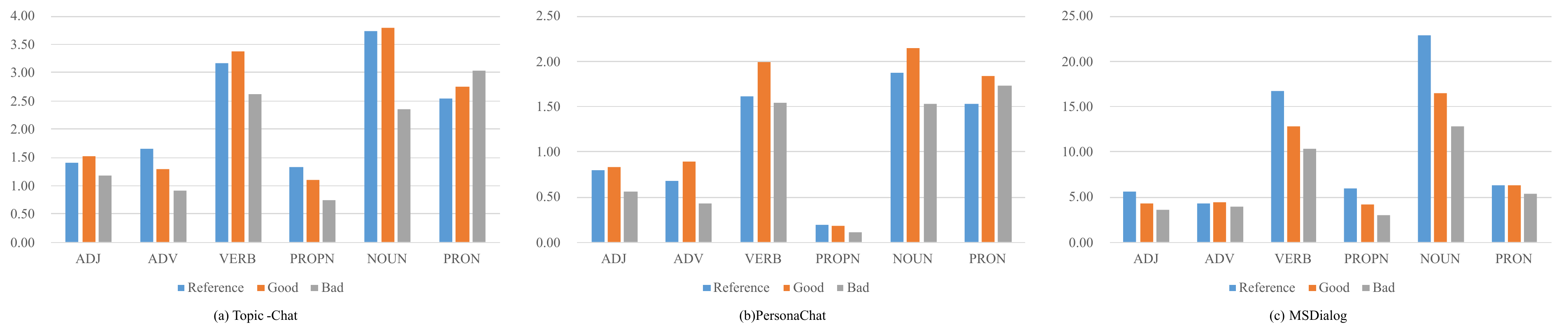}
    \vspace{-0.2in}
    \caption{The number distribution of selected POS tags in references, good responses, and bad responses. The x-axis denotes the select POS tags, and the y-axis denotes the average number of the POS tag in the corresponding dataset.}
    \label{fig:distribtion_pos_tags}
    \vspace{-0.1in}
\end{figure*}

Figure ~\ref{fig:distribtion_pos_tags} shows the average number of POS labels in each response, including references, for `good' and `bad' candidates (defined in \S\ref{sec:datasets}). 
%\lzy{
Although the distribution of POS tag words in those selected collections are different, some common trends can be observed across them: 
%1) 
the number of POS tag words found in the `good' candidates is closer to the number found in the references, in comparison to `bad' candidates. Most of the selected POS words (e.g., `ADJ', `ADV', `PROPN', `NOUN') follow this trend in all collections.\footnote{Note that `VERB' and `PRON' in PersonaChat are two exceptions, due to the unique characteristics of this collection.}
%This is because that the topics in PersonaChat concentrate on personal characteristics and the variance of the numbers of human-related words might be larger than those in the other two collections. Given that these two kinds of POS words remain the same trend in Topic-Chat and MSDialog, we believe this observation is widely suitable for most of the collection.} 
%2) The ways for POS words in `good' candidates to approximate the references are different. For example, it can be observed that the `NOUN' number in `good' candidates is greater than the number in reference in both Topic-Chat and PersonaChat, while the number of this type of words in `good' candidates is much less than that in reference in MSDialog.  }
%It can be observed that the distribution of POS tag words in selected collections are very different. For example, the number of PRON tag words of the references is much greater than the number of PROPN tags for both Topic-Chat and PersonaChat, while the number of these two types of words (i.e., PRON and PROPN) is more similar in the MSDialog corpus. However, it is worth noting that `good' quality candidates always have a similar number of POS tag words as the references, while `bad candidates' usually do not. %which means that the number of POS tag words in `good' candidates is generally closer to the number found in references.%, in comparison to `bad' candidates. 
In addition, the similarity between candidates and references in terms of POS tag words might be a useful signal to distinguish the quality of responses. 
%\lzy{Here we also notice that not all POS tag words follow this trend, such as VERB and PRON in PersonaChat. It is not surprising, however, since the topics in PersonaChat concentrate on personal characteristics and the variance of the numbers of human related words might be larger than those in the other two collections.} 
Based on the observed trend, we summarize two basic assumptions in this study:
\begin{itemize}
\item \noindent \textbf{Assumption 1}: The response of good quality should contain similar POS words as the reference response; 
\item \noindent \textbf{Assumption 2}: The difference in the distribution of POS tags between the candidate and reference responses is useful for measuring the relevance of the candidate response.
\end{itemize}

%A good quality conversational search response will have a similar distribution of POS tag words to the reference. The difference in POS tag word distribution, between candidate responses and references, may improve the evaluation of responses. 

%% file: content/3_3_incorporation_method.tex
To systematically analyze the effect of POS labelling, we firstly use simple methods to extend existing metrics with POS tags. We adopt two adaptive approaches for incorporating POS tag information: %(shown in Table ~\ref{tab:summary_incoporation}).

%\begin{itemize}
    %\item 
\noindent \textbf{POS Word Extraction (PWE)} - This is an intuitive way that only considers the words with specific POS tags. We extract all the words within POS tag sets (i.e., the specific tags which we choose) and filter other words (Table ~\ref{tab:example_pos_tag}). We then combine these POS words into a new sentence, which is used as an input to existing metrics.
    
\noindent \textbf{POS Tag Linear Combination (PTLC)} - We further consider the overlap of POS tag distribution on the basis of PWE. Firstly, we extract all the POS words and combine them into a new sentence like PWE. Then all the corresponding POS tags of these words are also extracted (shown in Table ~\ref{tab:example_pos_tag}). We then use different strategies to combine POS tag words and POS tags: 1) For hard matching metrics (i.e., BLEU), we directly combine POS tag words and POS tags together and put them into one new sentence as the metric input; 2) For synonym-based metrics (e.g., METEOR) and word embedding-based metrics (e.g., EA), we separately calculate the similarity of POS words and POS tags since POS tags do not have synonyms or embedding vectors. The original embedding-based metrics are used for calculating the similarity score of the extracted POS word text. We denote this score as the \textit{POS text score}. The overlap score of POS tags is computed by word overlapping-based metrics. In our paper, we use BLEU1 to compute the overlapping rate of the POS tag sequence. The score of these POS tags is denoted as \textit{POS tag score}. Finally, the POS text score and the POS tag score are linearly combined (added).
%, resulting in the ultimate score of the PTLC method.
    
%\end{itemize}

\begin{comment}
\begin{table}[] \footnotesize
\begin{tabular}{l|p{4cm}|p{1.5cm}}
\hline
Method & Description & Input \\ \hline
PWE & Only the words with the specific POS tags & POS words \\ \hline
PTLC & The similarity of POS words and the overlapping of POS tags & POS words + POS tags \\ \hline
\end{tabular}
\caption{A summary of POS-aware Adapted Metrics.}
\vspace{-0.1in}
\label{tab:summary_incoporation}
\end{table}
\end{comment}

\begin{table}[]%\small
\footnotesize
\caption{An example of POS words and POS tags when the tag set is <NOUN, VERB>. }
\vspace{-0.1in}
\begin{tabular}{p{1.5cm}|p{6cm}}
\hline
Orignal & it is from our evolution when land animals had both gills and lungs \\ \hline
POS words & <is, evolution, land, animals, had, gills, lungs> \\ \hline
POS tags & <VERB, NOUN, NOUN, NOUN, VERB, NOUN, NOUN> \\ \hline
POS words + POS tags &  \begin{tabular}[c]{p{6cm}} <is, evolution, land, animals, had, gills, lungs, \\ VERB, NOUN, NOUN, NOUN, VERB, NOUN, NOUN>  \end{tabular} \\ \hline
\end{tabular}
\label{tab:example_pos_tag}
\vspace{-0.2in}
\end{table}

Since POS tags generally indicate the grammatical role of a word in a sentence, different types of POS words may contain different information. Therefore, selected POS tags play a vital role in our proposed methods. To comprehensively analyze the effect of POS selection, we list all POS tag combinations that are adopted in the experiment in Table ~\ref{tab:pos_tag_set}. Here a POS tag set means we only select the words with the POS tags in the tag set. For example, `VERB' means we only extract the `VERB' tag words from the original sentence. It is worth noting that we consider `PROPN' and `NOUN' together since both tags are nominal attributes.

\begin{table}[t] %\small
\footnotesize
\caption{The POS tag combinations in the experiment. POS tag set means only the words with the corresponding POS tags (in the POS set) are selected for the evaluation. }
\vspace{-0.1in}
\begin{tabular}{ll}
\hline
\multicolumn{2}{c}{POS tag set} \\ \hline
\multicolumn{1}{l|}{ADJ} & ADJ + VERB + PROPN + NOUN \\
\multicolumn{1}{l|}{ADV} & ADJ + PROPN + NOUN + PRON \\
\multicolumn{1}{l|}{VERB} & ADV + VERB + PROPN + NOUN \\
\multicolumn{1}{l|}{PRON} & ADV + ADJ + PROPN + NOUN \\
\multicolumn{1}{l|}{PROPN + NOUN} & ADV + PROPN + NOUN + PRON \\
\multicolumn{1}{l|}{ADV + VERB} & VERB + PROPN + NOUN + PRON \\
\multicolumn{1}{l|}{VERB + PROPN + NOUN} & ADJ + ADV + VERB + PROPN + NOUN \\
\multicolumn{1}{l|}{PROPN + NOUN + PRON} & \multirow{2}{*}{\begin{tabular}[c]{@{}l@{}}ADJ + ADV + VERB + PROPN + NOUN\\ + PRON\end{tabular}} \\
\multicolumn{1}{l|}{ADJ + PROPN + NOUN} &  \\ \hline
\end{tabular}

\label{tab:pos_tag_set}
\vspace{-0.1in}
\end{table}

%% file: content/3_4_posscore_metric.tex
In this paper, our goal is to propose simple yet effective POS-based metrics for conversational search. The basic idea of our metrics is to increase the importance of selected POS words and give more weight to the POS similarity scores if the POS distribution is similar to the reference. 

Given a reference response $r = <r_1, r_2, r_3,...,r_j>$ and a candidate response $\hat{r} = <\hat{r}_1, \hat{r}_2, \hat{r}_3,..., \hat{r}_k>$, we use word embeddings to present the tokens. Then each response is split into two group: POS words sequence and Non-POS words sequence. The POS word sequence only contains the words with the selected POS tags, while the Non-POS sequence is the remaining words of the response. Thus, a reference is split as a POS word sequence $r_p=<r_{p1}, r_{p2}, ..., r_{pm}>$ and a Non-POS word sequence $r_q=<r_{q1}, r_{q2}, ..., r_{qm'}>$, and a candidate response is grouped as a POS word sequence $\hat{r}_p = <\hat{r}_{p1}, \hat{r}_{p2}, ..., \hat{r}_{pn}>$ and a Non-POS word sequence $\hat{r}_q = <\hat{r}_{q1}, \hat{r}_{q2}, ..., \hat{r}_{qn'}>$. The POSSCORE could be calculated as below shown in Equation \ref{equ:posscore}:

\vspace{-0.1in}
\begin{equation}
    POSSCORE(r,\hat{r})= w * S(r_p, \hat{r}_p) + S(r_q, \hat{r}_q)
    \label{equ:posscore}
\end{equation}
where $w$ means the weight function of POS tag rewarding, and $S(x, \hat{x})$ is the cosine similarity of the average embedding between sentence $x$ and $\hat{x}$ (defined in Equation \ref{equ:cosinesimilarity}).

\vspace{-0.1in}
\begin{equation}
    S(x, \hat{x})=cosine(\bar{E}(x),\bar{E}(\hat{x}))
    \label{equ:cosinesimilarity}
\end{equation}
POSSCORE should consider both the quality of POS word content and the difference in the distribution of POS tags between the references and candidate responses. Since the similarity score $S(r_p, \hat{r}_p)$ can capture the content-level similarity of POS words, the design of the weight function $w$ should consider the distribution difference. Therefore, the weight function needs to meet these requirements: 1) If the number of POS words in a candidate is less than that in the reference, the gain from POS words could be small and $w$ could reduce the importance of the POS similarity part; 2) If the number of POS distribution is the same, $w$ could be 1 and keep the original POS similarity scores; 3) If the number of POS words is larger than that in references, POS word part could be important and $w$ could increase the gain of POS similarity part. Therefore, following these criteria, the weight function $w$ is defined as Equation \ref{equ:weight_fun}. This weight function is inspired by the penalty function of BLEU metrics \cite{papineni2002bleu}. The range of $w$ is $0<w<e$.
\vspace{-0.1in}
\begin{equation}
    w=\exp (1 - \frac{n_r}{n_{\hat{r}}})
    \label{equ:weight_fun}
\end{equation}
where $n_r$ is the percentage of POS words in references (i.e., the number of selected POS words in reference $r$ divided by the length of reference $r$), and $n_{\hat{r}}$ is the percentage of POS words in candidate responses (i.e., the number of selected POS words in candidate $\hat{r}$ divided by the length of candidate $\hat{r}$). This weight function entails three different scenarios:

\begin{itemize}[nosep, wide=0pt]
    \item if $n_r > n_{\hat{r}}$, then w < 1. This means candidate responses do not have enough POS tag words by comparing to the references. In other words, the candidates may lack the necessary information expressed by the POS words in the reference response. Therefore, the gain from POS similarity scores should decrease.
    \item if $n_r = n_{\hat{r}}$, then w = 1. This means the POS distribution of candidate responses might be similar to the references. There is no extra gain for the POS word part.
    \item if $n_r < n_{\hat{r}}$, then w > 1. This means the number of the POS words in the candidate response is more than that in the reference and more likely to cover the POS information of references. In other words, the gain from POS word parts is more important than the rest of a candidate response. Therefore, $w$ increases the importance of the POS part and give more weight to the POS similarity scores.
\end{itemize}
%\todo{(Minor) It is not clear why we do not adopt percentage (i.e. normalized version). Here, a very strong assumption was made that the perfect answer would maintain similar length.}

Thus, we simply use the difference of POS word percentage between references and candidates to capture the distribution of POS information and dynamically control the importance of the POS similarity part. 
In order to find the influence of selected POS tags on the performance of POSSCORE, we also test different POS tag combinations as shown in Table \ref{tab:pos_tag_set}.

%% file: content/4_0_experimental_settings.tex
This section describes the key elements of our experimental setup, including dataset selection and pre-processing (\S\ref{sec:datasets}), baselines (\S\ref{sec:baseline}), and metric evaluation methods (\S\ref{sec:metric_evaluation}).

\subsection{Datasets} \label{sec:datasets}
\input{content/4_1_datasets.tex}

\subsection{Baseline Metrics} \label{sec:baseline}
\input{content/4_2_baseline_metrics.tex}

\subsection{Evaluating Automatic Evaluation Metrics} \label{sec:metric_evaluation}
\input{content/4_3_metric_evaluation.tex}

%% file: content/4_1_datasets.tex
In view of the complex interactions involved in conversational search, the datasets used for metric evaluation should meet three criteria: 1) Each interactive dialogue should have specific search intents since the existence of search intents is one of the important features in conversational search \cite{anand2020conversational}; 2) To refrain from collecting human annotations ourselves and reduce annotation bias, it is better that the open datasets contain human annotations that enable us to align metrics to the gold standard; 3) The dialogues in the datasets consist of multiple-round interactions so that the test environment can be closer to real conversational search scenario. After a comprehensive survey of existing datasets \cite{cohen2018wikipassageqa, qu2018analyzing, yang2015wikiqa, rajpurkar2016squad, li2017dailydialog, dinan2018wizard, lowe2015ubuntu, gopalakrishnan2019topical, zhang2018personalizing, ye2021multiwoz}, we chose three datasets that met the criteria: Topic-Chat \cite{gopalakrishnan2019topical}, PersonaChat \cite{zhang2018personalizing}, and MSDialog \cite{qu2018analyzing}.
%\begin{itemize}
%\item 
\noindent \textbf{Topic-Chat (TC) } \cite{gopalakrishnan2019topical} is a large collection of knowledge-grounded human-to-human conversations that consists of 11,319 dialogues with 8 broad topics. Each conversation has a specific topic and each utterance in the conversation is rated on a 5-point scale of quality.  
%\item 
\noindent \textbf{PersonaChat (PC) } \cite{zhang2018personalizing} is a dataset of human-to-human persona-conditioned open-domain conversations that contain 10,907 dialogues with personal topics. Each partner is asked to act as a persona to converse with each other.
%\item 
\noindent \textbf{MSDialog } \cite{qu2018analyzing} is a large-scale dialogue corpus of question answering interactions between customers and a help desk from an online forum on Microsoft products. This dataset consists of more than 2,000 multi-round information-seeking conversations with 10,000 utterances. 
%\end{itemize}

\noindent \textbf{Preprocessing of datasets.} Given that we adopt predictive power (described in \S\ref{sec:metric_evaluation}) to examine our metric, each conversation in above corpora needs to be converted to <question, reference, response1, response2> evaluation sets, which means one question can have one reference response (i.e., ground truth) and two candidate responses with different qualities, namely,  a `good' candidate response and a `bad' candidate response. Note that these two candidate responses answer the same question. Given the dataset differences, we adopt different strategies to preprocess them.

\begin{figure}[t]
    \centering
    \includegraphics[width=5cm]{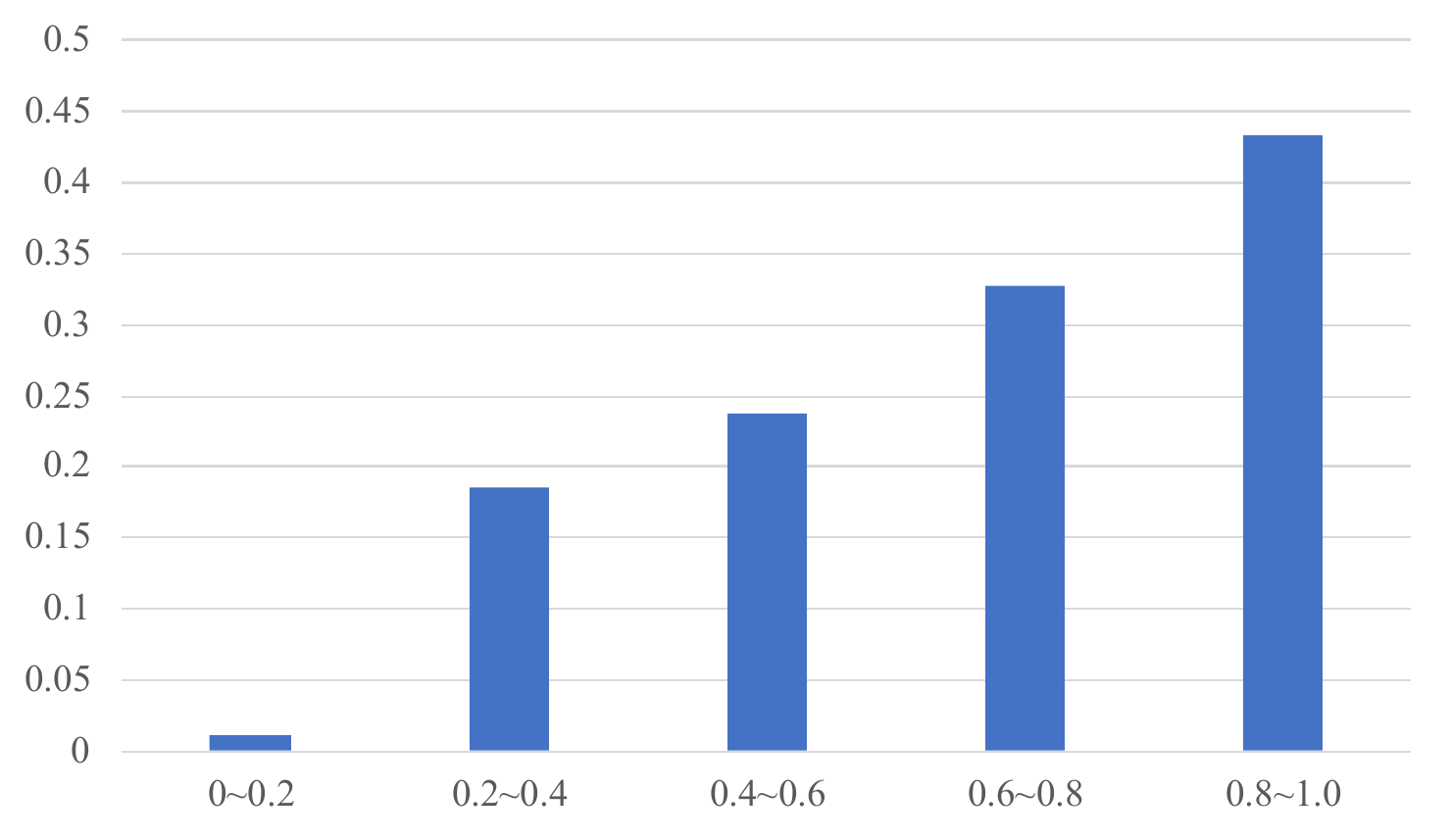}
    \vspace{-0.1in}
    \caption{The distribution of ground-truth answers in the voted responses. The x-axis denotes the normalized vote number, and the y-axis represents the proportion of the ground truth answers.}
    \label{fig:vote_gt}
    \vspace{-0.25in}
\end{figure}

In the original TC and PC corpora, a question only has one response, which is inadequate for our metric evaluation method. Therefore, we refer to the recent work by  \citet{mehri2020usr} and use their publicly released collections\footnote{The collections (i.e., TC and PC) are available at \url{http://shikib.com/usr}}, which are built on TC and PC corpora. \citet{mehri2020usr} conduct a human quality annotation of human-to-machine and human-to-human responses for both TC and PC to study the efficiency of their USR metric. In their collections, human annotation is carried out on sixty dialogue contexts. For each context, there are three or four system outputs (obtained from different generative models), one newly-written human response, and one ground-truth response. All the responses are labelled by three annotators with scores from six dimensions: Understandable (0-1), Natural (1-3), Maintains Context (1-3), Interesting (1-3), Uses Knowledge (0-1), and Overall Quality (1-5). In our study, we only consider the `Overall Quality' scores because this score intuitively reflects human preference. The average value of `Overall Quality' scores from three annotators is regarded as the final score for each response. Thus, all the system outputs and human response could be grouped into different <question, reference, response1, response2> sets in line with the final score for each context. If a response has a higher score than the other one, this response is deemed to be the `good' response in this set. It is worth noting that we do not put the responses with the same final scores into one set, which means candidates in one set should have different annotation scores. Finally, we obtain 550 sets for TC and 328 sets for PC (shown in Table ~\ref{tab:dataset}).

\begin{table}[t] \footnotesize
\caption{The number of evaluation sets in each collection.}
\vspace{-0.1in}
\begin{tabular}{c|c}
\hline
Corpus Name & Evaluation Sets \\ \hline
Topic-Chat(TC) & 550 \\
PersonaChat(PC) & 328 \\
MSDialog & 3,000 \\ \hline
\end{tabular}

\label{tab:dataset}
\vspace{-0.2in}
\end{table}

Unlike TC and PC with explicit human annotation scores, MSDialog collects users' judgements by using a variety of human feedback labels \footnote{MSDialog collection is available at \url{https://ciir.cs.umass.edu/downloads/msdialog/}}. In this paper, we use two types of human labels, namely `vote' and `is\_answer' to obtain evaluation sets. The tag `$vote$' represents the number of `helpful' votes for the answer from the community. If users agree with the response and think this answer may be helpful for this question, users can give one vote to this response. Note that users are not allowed to vote the same response more than once. The tag `$is\_answer$' is a binary tag, which indicates whether this answer is selected as the best answer in the dialogue session. Especially, this tag is often annotated by the user who posted the initial question and started the dialogue. Given that the `is\_answer' annotation can indicate that the response solves the issue from the questioners' perspective, we deem the $is\_answer$ responses to be the reference responses (ground truth).

To shed light on the connection between the `$vote$' and `$is\_answer$' tag, we further calculate the proportion of the ground truth against the voted responses. All the `$vote$' values are normalized by the maximum `$vote$' of the same dialogue in order to reduce the trendy questions' bias. Fig ~\ref{fig:vote_gt} shows the distribution of ground-truth answers in the voted responses. It is observed that the proportion of ground-truth answers grows steadily with the increase of the normalized vote scores, which means a response is more likely to be annotated as a relevant answer if this response has more votes in the same conversation session. It is worth noting that the proportion is 43.26\% when the normalized value is 1 (i.e., the response has the most votes in the dialogue). Therefore, we use the tag `$vote$' as an indicator to simulate the different human judgements. If a response has a higher `$vote$' value, it is regarded as the `good' response in a set. Finally, we randomly select \textasciitilde3,000 evaluation sets to test our metric.

%% file: content/4_2_baseline_metrics.tex
Since POS is a token-level label, using word-level baseline metrics may allow us to easily extend these metrics with POS labelling (shown in \S\ref{sec:incorporation_method}) and further find out the effect of part of speech in the evaluation process. Therefore, in this paper, three types of metrics were chosen (shown in Table ~\ref{tab:selection_metrics}):

%\begin{itemize}
%\item 
\noindent {\bf Word overlap-based metrics.} The basic idea of this type of metric is to count the number of words that co-occur in both the candidate responses and the ground truth. Here we choose BLEU \cite{papineni2002bleu} and METEOR  \cite{banerjee2005meteor}\footnote{Here we use the default settings of NLTK module, where $\alpha=0.9$, $\beta=3$, and $\gamma=0.5$. The module link is \url{https://www.nltk.org/_modules/nltk/translate/meteor_score.html}} as the baselines, which are popular metrics used in the evaluation of open-domain dialogue. 
    
    %\item 
\noindent {\bf Word embedding-based metrics.} The basic idea of these metrics is to using embedding information to connect between words that are semantically similar and evaluate the similarity between candidate sentences based on these word embedding vectors. Embedding Average (EA) \cite{foltz1998measurement, landauer1997solution, mitchell2008vector, serban2017hierarchical}, Embedding Extrema \cite{forgues2014bootstrapping} and Embedding Greedy \cite{rus2012comparison} are common representatives of this type of metrics. Since the performance of these three metrics are very close \cite{liu2016not}, we choose EA\footnote{We choose Fasttext embedding as the default embedding for EA and POSSCORE. } as the baseline of embedding-base metrics. Additionally, BERT-Score \cite{bert-score} is one latest proposed metrics, which has been demonstrated to have a high correlation with human judgments. Therefore, BERT-Score\footnote{Here we use the default settings of BERT-Score. The details of the settings are shown in \url{https://github.com/Tiiiger/bert_score} } which is also selected for a baseline.
    
    %\item 
    \noindent {\bf Learning-based metrics.} This type of metric always consists of one or more training models, such as ADEM \cite{lowe2017towards}, RUBER \cite{tao2018ruber}, PONE \cite{lan2020pone}, and BERT-RUBER \cite{ghazarian2019better}. Following previous work \cite{ghazarian2019better, liu2021meta}, we select BERT-RUBER as the sole representative learning-based metric in this study given its superior performance. Since the performance of learning-based models could be influenced by the pre-prepared training dataset \cite{liu2021meta}, we train and tune the model based on the specific dataset we use.

%\end{itemize}
%
%\todo{argue why we do not use learning based metrics as our baselines}
%Therefore, we consider most word overlap-based metrics and word embedding-based metrics.

\begin{table}[t] %\small
\footnotesize
\caption{The selected automatic evaluation metrics.}
\vspace{-0.1in}
\begin{tabular}{c|c}
\hline
Metric Category & Metric \\ \hline
Word overlap-based & BLEU1-4, METEOR \\
Word embedding-based & Emebedding Average, BERT-Score \\ 
Learning-based & BERT-RUBER \\ \hline
\end{tabular}
\label{tab:selection_metrics}
\vspace{-0.2in}
\end{table}

%% file: content/4_3_metric_evaluation.tex
Metric fidelity is the key concern in our metric design. Fidelity reflects the ability of a metric to measure what it intends to measure and agree with ultimate user preferences. Many recent studies \cite{mehri2020usr, yuma2020ubleu, sinha2020learning} use \textit{spearman's correlation coefficients} and \textit{Pearson's correlation coefficients} to test the correlation between metric judgements and human annotations. However, given that there are many ties among human annotation scores, the correlation score could be very close and it may be difficult to distinguish the performance of candidate metrics. Therefore, in our experiment, we adopt predictive power \cite{sanderson2010user} to capture the extent of a given evaluation metric's ability to predict a user's preference.
Predictive power measures the ability of metrics to describe the agreement between metrics and user preferences~\cite{sanderson2010user}. The basic idea of predictive power is that if an evaluation metric agrees with the user's preference between two outputs, then that is a correct prediction \cite{sakai2013metrics}. The higher the predictive power score is, the more similar to the human judgements the metric is. We use predictive power to examine the similarity between metrics and human judgements in conversational search (as shown in Algorithm ~\ref{alg:pdalg}).

\begin{algorithm}[t] %\footnotesize
\scriptsize
\caption{Computing predictive power.} 
\label{alg:pdalg}

Total=0;Correct=0\;
\For{d = 1 to N \tcp{for each dialogue}}{
\ForEach{pair of responses($r_1$, $r_2$) in dialogue d}{
Total++\;
$\delta X=X(r_1)-X(r_2)$\; \tcp{$X$ is the metric score of response $r_j$}
$\delta X^*=X^*(r_1)-X^*(r_2)$\; \tcp{$X^*$ is the judgement score of response $r_j$, such as votes}
\If{($(\delta X \times \delta X^*) > 0 $) \tcp{$X$ and $X^*$ positively agree}}{
    Correct++\;
}
}
}
PredictivePower = Correct/Total\;
\end{algorithm}

Here, we summarize the overall process of metric evaluation as follows: 1) We first collect all the <question, reference, response1, response2> sets from the three datasets. 2) Candidate metrics are adopted to calculate the scores for both response1 and response2. 3) After that, we use predictive power to examine the coherence between metric judgements and human preference. 4) Finally, different candidate metrics are compared on their predictive power. 

%% file: content/5_0_experiment_results.tex
\begin{table*}[t]
\footnotesize
    \centering
    \caption{The predictive power results of PWE methods. Baselines are calculated on the original sentences. The two-sided t-test is performed to detect any significant difference between proposed methods and baselines. * and ** represent significant value $p<0.05$ and $p<0.01$. The block colour represents the power scores' change of direction compared with the baseline in the same column (red denotes increment and blue represents decrements; the brightness of the colour indicates the change magnitude). }
    \vspace{-0.1in}
    \includegraphics[width=16cm]{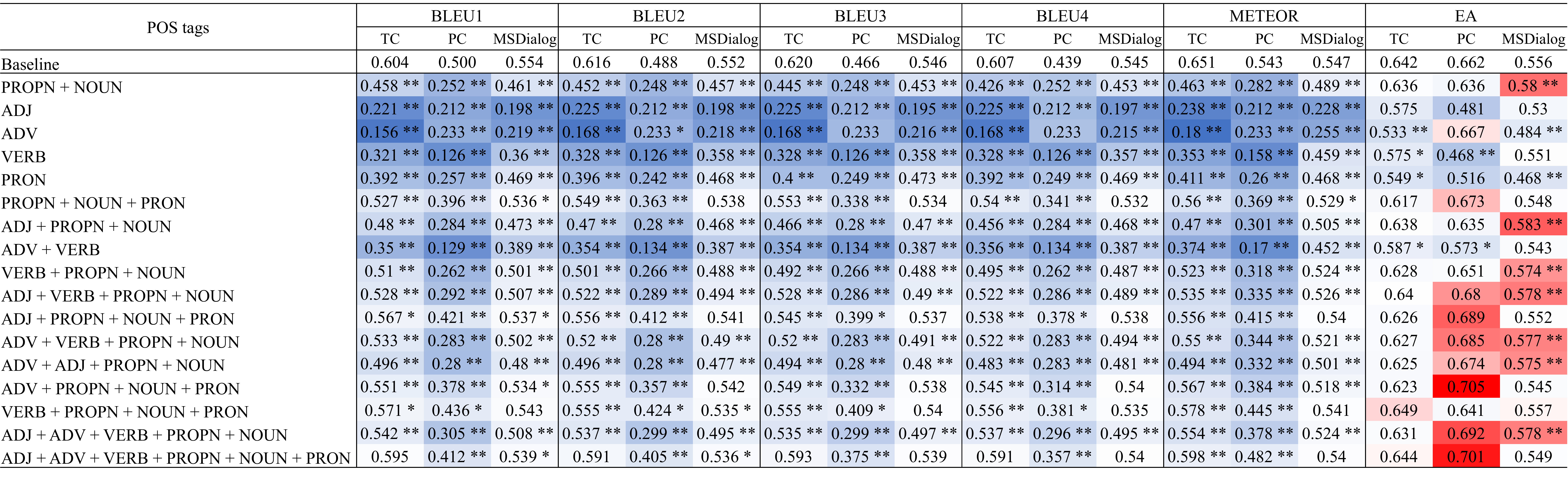}
    \label{tab:result_PWE}
    \vspace{-0.1in}
\end{table*}

This section presents the experimental results for both POS-aware Adapted measures (\S\ref{sec:result_incor}) and POSSCORE metrics (\S\ref{sec:result_posscore}). We systematically compare the performance of different POS combinations. We also discuss the effect of response length on POSSCORE (\S\ref{sec:result_effect_length}), the correlation between metrics (\S\ref{sec:result_correlation}), and a case study (\S\ref{sec:case_study}).

\subsection{POS-aware Adapted Metrics}
\label{sec:result_incor}
\input{content/5_1_incorporation_method_results.tex}

\subsection{POSSCORE}
\label{sec:result_posscore}
\input{content/5_2_posscore_results.tex}

%\subsection{Case study}
%\label{sec:case_study}
%\input{content/5_3_case_study.tex}

\subsection{Effect of Response Length}
\label{sec:result_effect_length}

\input{content/5_4_effect_of_response_length.tex}

\subsection{Correlation Analysis}
\label{sec:result_correlation}
\input{content/5_3_correlation_analysis.tex}

\subsection{Case Study}
\label{sec:case_study}
\input{content/5_5_case_study.tex}

%% file: content/5_1_incorporation_method_results.tex
\begin{table*}[]
    \centering
    
    \caption{The predictive power results of PTLC methods. Baselines are calculated on the original sentences without POS tags. The same annotation strategy of Table \ref{tab:result_PWE} is utilized for this table.}
    %The two-sided t-test is performed to detect any significant difference between the proposed methods and baseline. * and ** represent significant value $p<0.05$ and $p<0.01$.  The block colour represents the power scores' change of direction compared with the baseline in the same column. (red denotes increment and blue represents decrement, and the brightness of the colour indicates the change magnitude).}
    \vspace{-0.1in}
    \includegraphics[width=16cm]{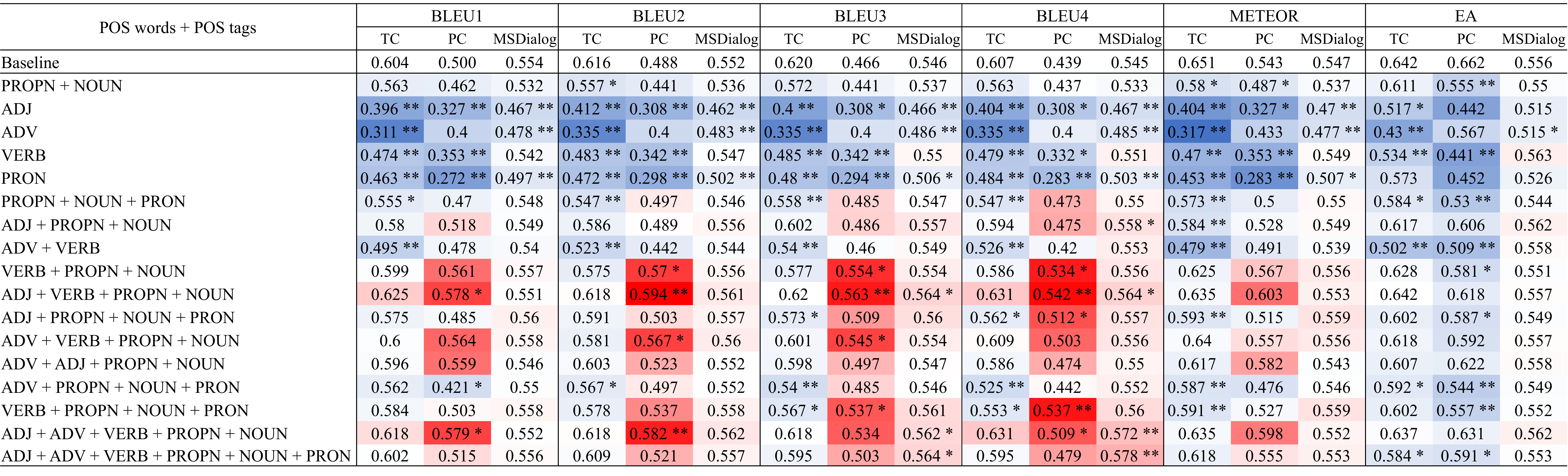}
    \label{tab:result_PTLC}
    \vspace{-0.1in}
\end{table*}

Table ~\ref{tab:result_PWE} shows the predictive power scores of PWE methods. The baseline is the results that are calculated with the original sentences. We use a two-sided T-test \cite{smucker2007comparison} to examine the difference between proposed methods and baselines.\footnote{Since we find the trends of the significant test with Bonferroni correction are similar to the original T-test, we present the original T-test results and set 0.05 and 0.01 as the significant thresholds in this paper.}
%\todo{describe an example of what the number means for one ``tab''.}
The scores in the table indicate the agreements between metrics and user preferences. For example, the baseline BLEU1 metric has 60.4\% (0.604) `correct' prediction, which agrees with users' preference (i.e., selecting `good' responses) within all the sessions in the TC collection.
We can observe that:

\begin{itemize}[nosep, wide=0pt]
    \item All PWE methods fail to defeat the original metrics based on word overlap (BLEU1-4 and METEOR), which means that only using the overlapping of POS words between candidates and references is inadequate for predicting users' preference. This is not surprising since the word reduction makes it more difficult for exact word matching.
    \item Although PWE with EA can perform better in PC, all the improvements are not significantly different to the baselines. This may be because of the small size of the PC collection. Meanwhile, We notice that many cases of PWE with EA significantly outperform the baselines in MSDialog, which means that PWE methods can provide a notable improvement on original EA metrics. 
    \item It is worth noting that PWE with EA has poor performance in the TC collection. Except for two cases (`VERB + PROPN + NOUN' and `ADJ + ADV + VERB + PROPN + NOUN'), other cases perform worse than the original EA. In other words, PWE methods are sensitive to the variance of collections. 
\end{itemize}

Table ~\ref{tab:result_PTLC} presents the predictive power results of PTLC methods. The baseline follows the same settings in PWE. %From this table, our findings are summarised as follows:
We can observe:

\begin{itemize}[nosep, wide=0pt]
    \item In general, the incorporation of POS tags significantly improves the performance of word overlap-based metrics. Comparing with the results of PWE with BLEU1-4 and METEOR, it can be observed that a large number of POS tag combinations tend to outperform the baselines after incorporating the POS tag sequence. These improvements, to some extent, reveal the positive effect of POS distribution in the evaluation process of word overlap-based metrics.
    
    \item The effects of promotion within PTLC methods are different across collections. For example, there are a number of significant improvement cases in both PC and MSdialog, whereas PTLC methods do not reach significant differences in TC.
    
    \item PTLC methods are also influenced by the original framework of the metrics. We can find BLEU and METEOR achieve more significant improvements after considering the POS information.
    
    \item The selected POS tags have a considerable influence on the performance of PTLC. Particularly, it can be observed that both `ADJ + VERB + PROPN + NOUN' and `ADJ + ADV + VERB + PROPN + NOUN' are robust and create the most significant improvement cases across all three collections.
    
    \item Comparing with the PWE results,  the predictive power of embedding-based metrics performs worse after incorporating POS tags - in most cases, they tend to perform poorer than the baselines. That means that rough incorporation of POS tags is inadequate for the improvement over the original embedding-based metrics. 
    
\end{itemize}
To sum up, we find roughly using POS words (i.e., PWE and PTLC methods) may be inadequate for high-quality evaluation. However, the performance of PWE methods could be improved significantly when considering word embedding-based metrics. PTLC results reveal the positive effects of using POS tag distribution in the evaluation process of overlap-based metrics. However, PTLC methods have limited improvements on embedding-based metrics.

%% file: content/5_2_posscore_results.tex
\begin{table}[t]
\footnotesize
    \centering
    \caption{The predictive power results of POSSCORE methods. Baselines are the original metrics' results with the raw response sentences. The two-sided t-test is performed to detect any significant difference between proposed methods and best candidate baselines. sig\_b means the significant test between POSSCORE and original metrics which achieve the best performance among all candidate baselines (e.g., METEOR in TC, and EA in PC and MSDialog ), and sig\_p mean the significant test between POSSCORE and the results of corresponding PWE methods. * and ** represent significant value $p<0.05$ and $p<0.01$. The block colour represents the power scores' change of direction compared with the `best' baseline in the same column (red denotes increment and blue represents decrement, and the brightness of the colour indicates the change magnitude).}
    \vspace{-0.1in}
    \includegraphics[width=8cm]{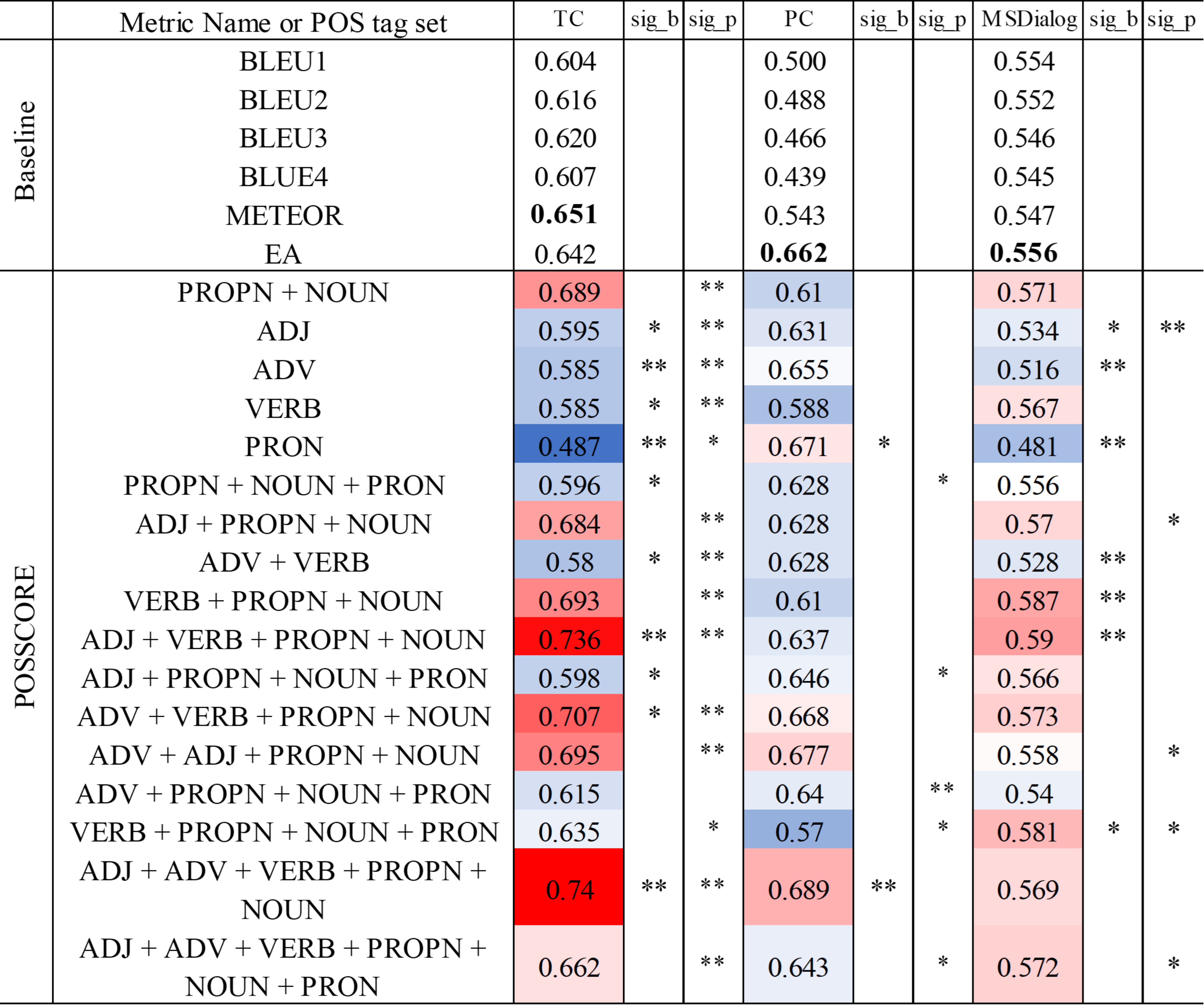}
    \label{tab:result_POSSCORE}
\vspace{-0.45in}
\end{table}

Both PWE and PTLC methods are simple extensions of existing metrics. Although PTLC has achieved significant improvement over the original metrics, these extension methods are not adequate to capture the connection of POS distributions between references and candidates. This motivated our new development in POS-based metrics: POSSCORE. Table ~\ref{tab:result_POSSCORE} presents the predictive power results for POSSCORE. In this experiment, we also adopt a two-sided T-test to examine significant differences. To comprehensively analyze our proposed metrics, we perform significance tests against both:
%\begin{itemize}
    %\item 
    (1) \emph{The best original baseline metric (sig\_b)}: for example, in TC, the original METEOR reaches the highest predictive power score and so METEOR is selected as the baseline of TC. In MSDialog, the original EA achieves the highest predictive power score and it is regarded as the baseline for MSDialog.
    %\item 
    (2) \emph{The PWE methods with the `best' original baseline metric (sig\_p)}: this is because we find PWE methods can achieve better performance than the original baseline (as discussed in \S\ref{sec:result_incor}). By doing so, we can identify any improvement of POSSCORE over the gains from PWE methods.   
%\end{itemize}

With the above settings, we can observe in Table~\ref{tab:result_POSSCORE} that:

\begin{itemize}[nosep, wide=0pt]
    \item Most configurations of POSSCORE can outperform the best performing baseline metrics. Especially, in TC, we can find that many POS tag sets can reach significant improvements against the original baseline. From this perspective, POSSCORE is better than PTLC methods since it can achieve significant improvements across all three collections. 
    \item Although a corpus may have an influence on the extent of performance improvement, POSCORE can robustly outperform the original baseline across all the selected corpora. It can be observed that although the improvement in MSDialog is smaller than that in TC and PC, POSSCORE is still able to achieve higher predictive power than the baselines.
    \item Comparing with the results of PWE, we can see POSSCORE is significantly better than PWE across all three datasets. 
    \item In terms of different POS combinations, it can be observed that `ADJ + ADV + VERB + PROPN + NOUN' can robustly achieve high predictive scores across three collections, especially with significant improvements in both TC and PC. Therefore, we recommend to choose \emph{`ADJ + ADV + VERB + PROPN + NOUN'} for POSSCORE. 
\end{itemize}

\begin{table}[t] %\small
\footnotesize
\caption{The predictive power comparison with BERT-Score and BERT-RUBER. Two-sided t-test is performed to detect any significant difference between POSScore and two state-of-the-art metrics. * and ** represent $p<0.05$ and $p<0.01$.}
\vspace{-0.1in}
\begin{tabular}{c|ccc}
\hline
           & TC             & PC             & MSDialog        \\ \hline
POSSCORE   & \textbf{0.740} & \textbf{0.689} & 0.569           \\ \hline
BERT-Score & 0.655**        & 0.607*         & 0.541*          \\
BERT-RUBER & 0.509**        & 0.561**        & \textbf{0.69**} \\ \hline
\end{tabular}
\vspace{-0.1in}
\label{tab:SOTA_metric}
\end{table}

Table ~\ref{tab:SOTA_metric} shows the comparison between POSSCORE and other more recent state-of-the-art metrics: BERT-Score \cite{bert-score} and BERT-RUBER \cite{ghazarian2019better}. Here we use `ADJ + ADV + VERB + PROPN + NOUN' to calculate POSSCORE as it generally performs the best (as shown in Table~\ref{tab:result_POSSCORE}). Since BERT-RUBER needs to pre-train unreferenced models, we split each dataset into training datasets (80\% of the whole datasets), develop datasets (10\% of the whole datasets), and test datasets (10\% of the whole datasets). Following previous work \cite{ghazarian2019better}, we also use 2 layers of the bidirectional gated recurrent unit with the 128-dimensional hidden unit and apply three layers for MLP (Multilayer Perceptron Network) with 256, 512 and 128-dimensional hidden units. Learning rate decay is applied when no improvement was observed on validation data for five consecutive epochs.

It can be observed that POSSCORE consistently outperforms BERT-Score and BERT-RUBER in TC and PC. All the improvements in these two datasets are significant. It is worth noting that BERT-RUBER outperforms the other two metrics in MSDialog, while it performs the worst in TC and PC. We found that the learning-based metric BERT-RUBER is quite sensitive to the training collection. By learning BERT-RUBER on five different random samples of the MSDialog dataset, we found its predictive power ranges from 0.66 to 0.71. Considering the poor performance of BERT-RUBER in TC and PC, this demonstrated that BERT-RUBER is not robust.
%\lzy{To provide insights on the variation of BERT-RUBER, We further perform BERT-RUBER five times on MSDialog. We notice that even we use the same training datasets, the BERT-RUBER still varies after unreference model training. The range of its predictive score is from 0.66 to 0.71 (Average is 0.697 and variance is 0.0005) in these five runs, which means BERT-RUBER is very sensitive. 
%Considering the poor performance of BERT-RUBER in TC and PC, BERT-RUBER is not as robust as our proposed metric, which can be significantly better than BERT-Score in all three datasets. In other words, 
Comparatively speaking, the performance of POSSCORE is more stable across different collections and is much easier to interpret.
%Therefore, the proposed POSSCORE has a stronger correlation with human preference than most of state-of-the-art metrics and the improvements are statistically significant across three different collections.
%\todo{This is not surprising because the small size of TC and PC is inadequate for training robust unreferenced models. When we adapt larger datasets, such as MSDialog, to train the models, the performance of the BERT-RUBER increase significantly.} Although POSSCORE are not able to outperform BERT-RUBER in MSDialog, we can find POSSCORE is still significantly better than BERT-Score in all three datasets. Therefore, the proposed POSSCORE has a stronger correlation with human preference than most of state-of-the-art metrics and the improvements are statistically significant across three different collections. 

%\todo{add correlation of metric results (perhaps a new subsection)}

%% file: content/5_4_effect_of_response_length.tex
\begin{table}[t] \footnotesize
%\vspace{-0.1in}
\caption{The predictive power results of modified test collection with longer bad candidate responses where their contents are duplicated. Thus, all the bad responses have doubled their lengths compared to the original bad responses.}
\vspace{-0.1in}
\begin{tabular}{c|ccc}
\hline
         & TC             & PC            & MSDialog       \\ \hline
BLEU2    & 0.718          & 0.652         & 0.535          \\
BLEU4    & 0.715          & 0.698         & 0.541          \\
METEOR   & 0.653          & 0.616         & 0.508          \\
EA       & 0.642          & 0.662         & 0.556          \\ \hline
POSSCORE & \textbf{0.756} & \textbf{0.71} & \textbf{0.587} \\ \hline
\end{tabular}
\label{tab:length_bias}
\vspace{-0.1in}
\end{table}

Since POSSCORE performs counting on the number of POS and Non-POS words and tags, we would like to understand the influence of response length on its performance. To examine the effect of length bias, we created a new dataset based on each existing dataset, in which we simply modify the low-quality responses to make them repeat the content twice to increase their length. Thus, all the low quality (bad) responses are twice as long as the original responses.
We compare POSSCORE against those best-performing metrics that were designed to be agnostic to the length (e.g.,~with length normalization), such as BLEU, METEOR and EA.  The hypothesis is that the performance of those length-agnostic metrics should not be significantly affected.
%\lzy{Since BLEU, METEOR and EA have their own special design to reduce the length influence, here we only choose the metrics which contain the design of length penalty as the baseline.} 
%Therefore, in this new dataset, we use the best performing BLEU2, BLEU4, METEOR, EA and POSSCORE to examine the influence of response length. 
Table ~\ref{tab:length_bias} presents the predictive power results of different metrics on the modified dataset. It can be observed that POSSCORE is not significantly affected by the length variation of the bad responses. All the results from POSSCORE are better than those selected baseline metrics. %Therefore, we can see the proposed POSSCORE is free from the length bias effects.

%% file: content/5_3_correlation_analysis.tex
\begin{figure}[t]
%\vspace{-0.1in}
\footnotesize
    \centering
    \includegraphics[width=5.1cm]{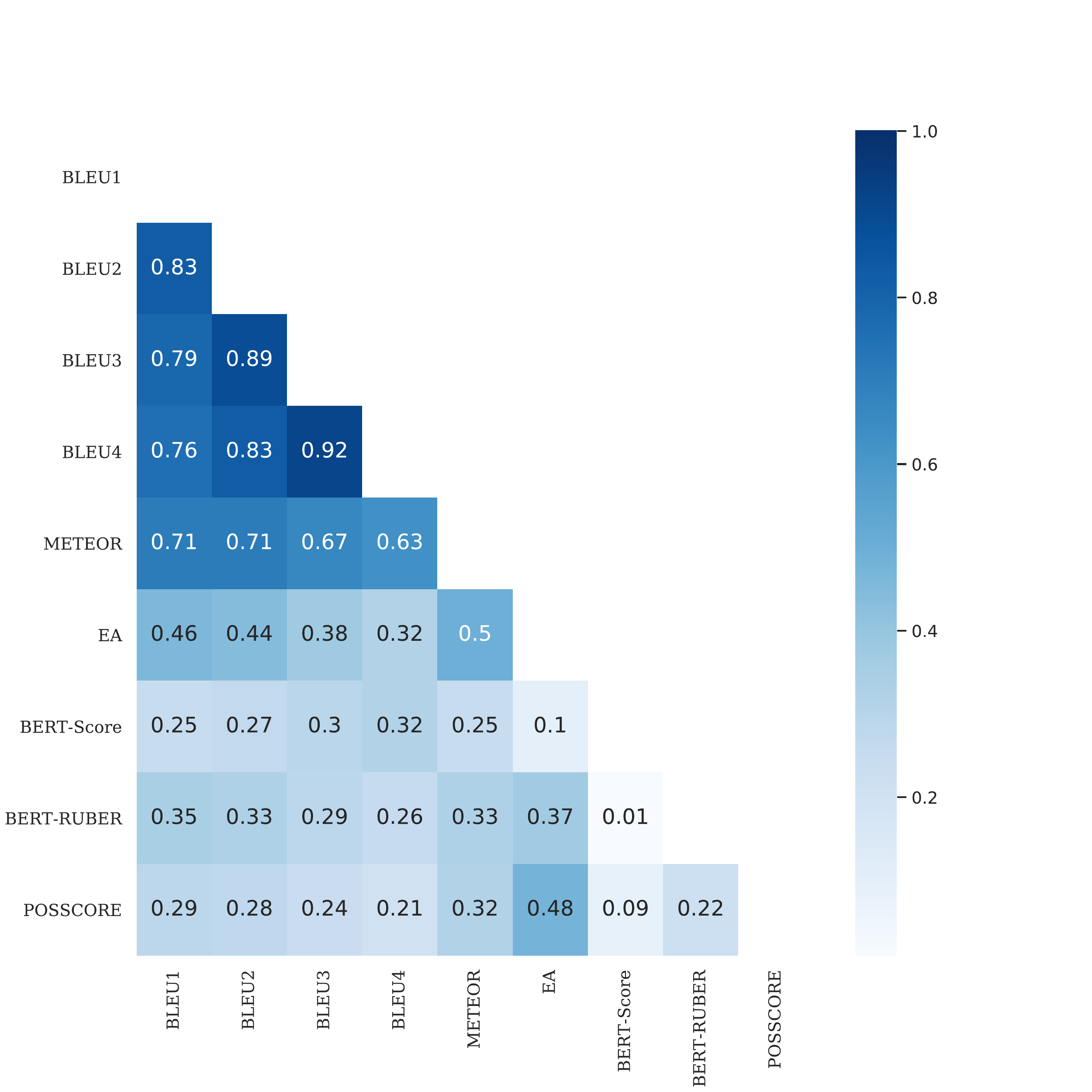}
    \vspace{-0.1in}
    \caption{The Kendall's Tau correlation between different metrics on MSDialog collection (similar trends are observed in the other two collections). }
    \label{fig:correlation}
    \vspace{-0.2in}
\end{figure}

%The previous sections compare the predictive power of different metrics. 
Following previous work \cite{sakai2008information, sakai2006evaluating, sakai2007reliability}, we compare the correlation between different metrics to analyze their relationships.
%further analyze the connection between metrics. }
%
%\todo{a more fine-grained analysis is required on what happens when two metrics disagree. See the case study section}
Figure ~\ref{fig:correlation} presents Kendall's Tau correlation ($r$) between each pair of metrics given the original references and system responses. %The POSSCORE is the version with `ADJ + ADV + VERB + PROPN + NOUN' setting. 
First of all, compared to correlations between traditional IR metrics based on relevance judgments of documents \cite{sakai2008information}, we find conversational search metrics are generally more weakly correlated with each other. This is due to the nature of those metrics, which leverage different ways of modelling similarity between candidate responses and references, as a proxy of relevance.
Secondly, not surprisingly, the metrics within the same category (Table \ref{tab:dataset}) are more strongly correlated with each other, whereas metrics across categories are only moderately correlated. For example, BLEU1-4 and METEOR are strongly correlated ($r$ > 0.6) while BERT-RUBER and BERT-Score are moderate ($r$ < 0.4). It is worth noting that the correlation of our proposed POSSCORE with other metrics are moderate ($r$ < 0.5), which means POSSCORE is measuring the responses substantially different from those metrics. Although POSSCORE is inspired by EA metrics, we can observe that the correlation between POSSCORE and EA is still less than 0.5. Therefore, this demonstrates that our proposed POSSCORE is a new metric that captures different aspects from those existing ones. 

%% file: content/5_5_case_study.tex
\begin{comment}

\begin{table}[t]\footnotesize
\caption{An example of failure of POSSCORE in the PersonaChat collection. The words with blue colour are the POS words that are recognized by POSSCORE.}
\vspace{-0.1in}
\begin{tabular}{c|l|c|c|c|c|c}
\hline
Reference & i \textcolor{blue}{love blue too}, i \textcolor{blue}{also enjoy mountain biking}. have you \textcolor{blue}{ever tried} it? & Human & BLEU4 & METEOR & BERT-Score & POSSCORE \\ \hline
Candidate1 & do you have any \textcolor{blue}{hobbies}? i \textcolor{blue}{enjoy} \textcolor{blue}{mountain biking}! & \textbf{4} & \textbf{0.067} & \textbf{0.379} & \textbf{0.880} & 1.437 \\ \hline
Candidate2 & i \textcolor{blue}{like pink} , i \textcolor{blue}{think blue} is \textcolor{blue}{too masculine color}. & 3 & 0.031 & 0.147 & 0.859 & \textbf{1.492} \\ \hline
\end{tabular}
\label{tab:fail_case}
\vspace{-0.1in}
\end{table}
\end{comment}

\begin{table}[t] \footnotesize
\caption{An example of failure case of POSSCORE in the PersonaChat collection. The words with blue colour are the POS words that are recognized by POSSCORE.}
\vspace{-0.1in}
\begin{tabular}{|c|l|c|c|c|c|}
\hline
Reference & \multicolumn{5}{l|}{I \textcolor{blue}{love blue too}. I \textcolor{blue}{also enjoy mountain biking}. Have you \textcolor{blue}{ever tried} it?} \\ \hline
Candidate1 & \multicolumn{5}{l|}{Do you have any \textcolor{blue}{hobbies}? I \textcolor{blue}{enjoy} \textcolor{blue}{mountain biking}!} \\ \hline
Candidate2 & \multicolumn{5}{l|}{I \textcolor{blue}{like pink} . I \textcolor{blue}{think blue} is \textcolor{blue}{too masculine color}.} \\ \hline
Evaluation & \multicolumn{1}{c|}{Human} & BLEU4 & METEOR & BERT-Score & POSSCORE \\ \hline
Candidate1 & \multicolumn{1}{c|}{\textbf{4}} & \textbf{0.067} & \textbf{0.379} & \textbf{0.880} & 1.437  \\ \hline
Candidate2 & \multicolumn{1}{c|}{3}  & 0.031 & 0.147 & 0.859 & \textbf{1.492}  \\ \hline
\end{tabular}
\label{tab:fail_case}
\vspace{-0.15in}
\end{table}

Despite that our proposed POSSCORE outperforms existing metrics, we present an example of its failure case in Table ~\ref{tab:fail_case} to demonstrate its limitations for potential further improvements. We notice that in this example, all the automatic metrics except for POSSCORE are consistent with human annotations. This failure might be due to that POSSCORE is more sensitive to the POS distribution, especially when the POS word embeddings are very close (e.g., 'blue' vs. 'pink' in Table ~\ref{tab:fail_case}). Although the length of candidate1 is longer than that of candidate2, the number of POS words in candidate2 (n=7) is closer to the number in the reference (n=9). Since POSSCORE treats all the selected POS words equal and provides bonus scores (controlled by $w$) to the response if the distribution of POS words is similar to the reference, POSSCORE is more likely to overestimate the quality of this kind of response. In other words, the performance of POSSCORE might be constrained by the number of POS words in the candidates (i.e.,~the performance drops when the candidates have fewer POS words than the references). It can also be observed that the different mechanism between POSSCORE and other metrics, where POSSCORE considers both the word embedding similarity and the POS distribution similarity.
%\todo{analysis of metric difference, the role of w, limitations}

%% file: content/6_conclusion.tex
In this paper, we systematically analyzed whether and how POS words and POS tags could be used to benefit the automatic evaluation of conversational search systems. Based on the analysis, we propose two simple approaches to incorporate POS information into existing evaluation metrics, and present POSSCORE as a new simple metric for evaluating conversational search based on the distribution of POS words and tags. Extensive experiments on three publicly available datasets show that POSSCORE achieves significantly better alignments with human preferences than baseline metrics. Our work sheds light on the effectiveness of leveraging syntactic information for conversational search evaluation.
In our future work, we plan to extend POSSCORE in two directions: (1) we currently treat all POS words within each syntactic group equally. However, we can observe in Table ~\ref{tab:result_PWE} that different POS groups play different roles. For example, the words with `NOUN' tags  have a stronger influence on the evaluation performance. 
(2) it has been shown that conversation context may contain useful explicit and implicit information for evaluation  \cite{lipani2021doing}, which we did not yet consider. We would like to propose metrics that are both context- and syntactic-aware.

%% file: sample-sigconf.bbl

\begin{thebibliography}{54}

%%% the Web address.

\ifx \showCODEN    \undefined \def \showCODEN     #1{\unskip}     \fi
\ifx \showDOI      \undefined \def \showDOI       #1{#1}\fi
\ifx \showISBNx    \undefined \def \showISBNx     #1{\unskip}     \fi
\ifx \showISBNxiii \undefined \def \showISBNxiii  #1{\unskip}     \fi
\ifx \showISSN     \undefined \def \showISSN      #1{\unskip}     \fi
\ifx \showLCCN     \undefined \def \showLCCN      #1{\unskip}     \fi
\ifx \shownote     \undefined \def \shownote      #1{#1}          \fi
\ifx \showarticletitle \undefined \def \showarticletitle #1{#1}   \fi
\ifx \showURL      \undefined \def \showURL       {\relax}        \fi
% invisible to TeX
\providecommand\bibfield[2]{#2}
\providecommand\bibinfo[2]{#2}
\providecommand\natexlab[1]{#1}
\providecommand\showeprint[2][]{arXiv:#2}

\bibitem[\protect\citeauthoryear{Anand, Cavedon, Joho, Sanderson, and
  Stein}{Anand et~al\mbox{.}}{2020}]%
        {anand2020conversational}
\bibfield{author}{\bibinfo{person}{Avishek Anand}, \bibinfo{person}{Lawrence
  Cavedon}, \bibinfo{person}{Hideo Joho}, \bibinfo{person}{Mark Sanderson},
  {and} \bibinfo{person}{Benno Stein}.} \bibinfo{year}{2020}\natexlab{}.
\newblock \showarticletitle{Conversational Search (Dagstuhl Seminar 19461)}. In
  \bibinfo{booktitle}{\emph{Dagstuhl Reports}}, Vol.~\bibinfo{volume}{9}.
  Schloss Dagstuhl-Leibniz-Zentrum f{\"u}r Informatik.
\newblock


\bibitem[\protect\citeauthoryear{Banerjee and Lavie}{Banerjee and
  Lavie}{2005}]%
        {banerjee2005meteor}
\bibfield{author}{\bibinfo{person}{Satanjeev Banerjee} {and}
  \bibinfo{person}{Alon Lavie}.} \bibinfo{year}{2005}\natexlab{}.
\newblock \showarticletitle{METEOR: An automatic metric for MT evaluation with
  improved correlation with human judgments}. In
  \bibinfo{booktitle}{\emph{Proceedings of the acl workshop on intrinsic and
  extrinsic evaluation measures for machine translation and/or summarization}}.
  \bibinfo{pages}{65--72}.
\newblock


\bibitem[\protect\citeauthoryear{Brill}{Brill}{1992}]%
        {brill1992simple}
\bibfield{author}{\bibinfo{person}{Eric Brill}.}
  \bibinfo{year}{1992}\natexlab{}.
\newblock \bibinfo{booktitle}{\emph{A simple rule-based part of speech
  tagger}}.
\newblock \bibinfo{type}{{T}echnical {R}eport}.
  \bibinfo{institution}{PENNSYLVANIA UNIV PHILADELPHIA DEPT OF COMPUTER AND
  INFORMATION SCIENCE}.
\newblock


\bibitem[\protect\citeauthoryear{Chen, Zhou, Liu, Zhang, and Ma}{Chen
  et~al\mbox{.}}{2017}]%
        {chen2017meta}
\bibfield{author}{\bibinfo{person}{Ye Chen}, \bibinfo{person}{Ke Zhou},
  \bibinfo{person}{Yiqun Liu}, \bibinfo{person}{Min Zhang}, {and}
  \bibinfo{person}{Shaoping Ma}.} \bibinfo{year}{2017}\natexlab{}.
\newblock \showarticletitle{Meta-evaluation of online and offline web search
  evaluation metrics}. In \bibinfo{booktitle}{\emph{Proceedings of the 40th
  International ACM SIGIR Conference on Research and Development in Information
  Retrieval}}. ACM, \bibinfo{pages}{15--24}.
\newblock


\bibitem[\protect\citeauthoryear{Clarke, Kolla, Cormack, Vechtomova, Ashkan,
  B{\"u}ttcher, and MacKinnon}{Clarke et~al\mbox{.}}{2008}]%
        {clarke2008novelty}
\bibfield{author}{\bibinfo{person}{Charles~LA Clarke},
  \bibinfo{person}{Maheedhar Kolla}, \bibinfo{person}{Gordon~V Cormack},
  \bibinfo{person}{Olga Vechtomova}, \bibinfo{person}{Azin Ashkan},
  \bibinfo{person}{Stefan B{\"u}ttcher}, {and} \bibinfo{person}{Ian
  MacKinnon}.} \bibinfo{year}{2008}\natexlab{}.
\newblock \showarticletitle{Novelty and diversity in information retrieval
  evaluation}. In \bibinfo{booktitle}{\emph{Proceedings of the 31st annual
  international ACM SIGIR conference on Research and development in information
  retrieval}}. \bibinfo{pages}{659--666}.
\newblock


\bibitem[\protect\citeauthoryear{Cohen, Yang, and Croft}{Cohen
  et~al\mbox{.}}{2018}]%
        {cohen2018wikipassageqa}
\bibfield{author}{\bibinfo{person}{Daniel Cohen}, \bibinfo{person}{Liu Yang},
  {and} \bibinfo{person}{W~Bruce Croft}.} \bibinfo{year}{2018}\natexlab{}.
\newblock \showarticletitle{Wikipassageqa: A benchmark collection for research
  on non-factoid answer passage retrieval}. In \bibinfo{booktitle}{\emph{The
  41st International ACM SIGIR Conference on Research \& Development in
  Information Retrieval}}. \bibinfo{pages}{1165--1168}.
\newblock


\bibitem[\protect\citeauthoryear{Dinan, Roller, Shuster, Fan, Auli, and
  Weston}{Dinan et~al\mbox{.}}{2018}]%
        {dinan2018wizard}
\bibfield{author}{\bibinfo{person}{Emily Dinan}, \bibinfo{person}{Stephen
  Roller}, \bibinfo{person}{Kurt Shuster}, \bibinfo{person}{Angela Fan},
  \bibinfo{person}{Michael Auli}, {and} \bibinfo{person}{Jason Weston}.}
  \bibinfo{year}{2018}\natexlab{}.
\newblock \showarticletitle{Wizard of wikipedia: Knowledge-powered
  conversational agents}.
\newblock \bibinfo{journal}{\emph{arXiv preprint arXiv:1811.01241}}
  (\bibinfo{year}{2018}).
\newblock


\bibitem[\protect\citeauthoryear{Foltz, Kintsch, and Landauer}{Foltz
  et~al\mbox{.}}{1998}]%
        {foltz1998measurement}
\bibfield{author}{\bibinfo{person}{Peter~W Foltz}, \bibinfo{person}{Walter
  Kintsch}, {and} \bibinfo{person}{Thomas~K Landauer}.}
  \bibinfo{year}{1998}\natexlab{}.
\newblock \showarticletitle{The measurement of textual coherence with latent
  semantic analysis}.
\newblock \bibinfo{journal}{\emph{Discourse processes}} \bibinfo{volume}{25},
  \bibinfo{number}{2-3} (\bibinfo{year}{1998}), \bibinfo{pages}{285--307}.
\newblock


\bibitem[\protect\citeauthoryear{Forgues, Pineau, Larchev{\^e}que, and
  Tremblay}{Forgues et~al\mbox{.}}{2014}]%
        {forgues2014bootstrapping}
\bibfield{author}{\bibinfo{person}{Gabriel Forgues}, \bibinfo{person}{Joelle
  Pineau}, \bibinfo{person}{Jean-Marie Larchev{\^e}que}, {and}
  \bibinfo{person}{R{\'e}al Tremblay}.} \bibinfo{year}{2014}\natexlab{}.
\newblock \showarticletitle{Bootstrapping dialog systems with word embeddings}.
  In \bibinfo{booktitle}{\emph{Nips, modern machine learning and natural
  language processing workshop}}, Vol.~\bibinfo{volume}{2}.
\newblock


\bibitem[\protect\citeauthoryear{Ghazarian, Wei, Galstyan, and Peng}{Ghazarian
  et~al\mbox{.}}{2019}]%
        {ghazarian2019better}
\bibfield{author}{\bibinfo{person}{Sarik Ghazarian}, \bibinfo{person}{Johnny
  Tian-Zheng Wei}, \bibinfo{person}{Aram Galstyan}, {and}
  \bibinfo{person}{Nanyun Peng}.} \bibinfo{year}{2019}\natexlab{}.
\newblock \showarticletitle{Better automatic evaluation of open-domain dialogue
  systems with contextualized embeddings}.
\newblock \bibinfo{journal}{\emph{arXiv preprint arXiv:1904.10635}}
  (\bibinfo{year}{2019}).
\newblock


\bibitem[\protect\citeauthoryear{Gopalakrishnan, Hedayatnia, Chen, Gottardi,
  Kwatra, Venkatesh, Gabriel, Hakkani-T{\"u}r, and AI}{Gopalakrishnan
  et~al\mbox{.}}{2019}]%
        {gopalakrishnan2019topical}
\bibfield{author}{\bibinfo{person}{Karthik Gopalakrishnan},
  \bibinfo{person}{Behnam Hedayatnia}, \bibinfo{person}{Qinglang Chen},
  \bibinfo{person}{Anna Gottardi}, \bibinfo{person}{Sanjeev Kwatra},
  \bibinfo{person}{Anu Venkatesh}, \bibinfo{person}{Raefer Gabriel},
  \bibinfo{person}{Dilek Hakkani-T{\"u}r}, {and} \bibinfo{person}{Amazon~Alexa
  AI}.} \bibinfo{year}{2019}\natexlab{}.
\newblock \showarticletitle{Topical-Chat: Towards Knowledge-Grounded
  Open-Domain Conversations.}. In \bibinfo{booktitle}{\emph{INTERSPEECH}}.
  \bibinfo{pages}{1891--1895}.
\newblock


\bibitem[\protect\citeauthoryear{Hashemi, Williams, El~Kholy, Zitouni, and
  Crook}{Hashemi et~al\mbox{.}}{2018}]%
        {hashemi2018measuring}
\bibfield{author}{\bibinfo{person}{Seyyed~Hadi Hashemi}, \bibinfo{person}{Kyle
  Williams}, \bibinfo{person}{Ahmed El~Kholy}, \bibinfo{person}{Imed Zitouni},
  {and} \bibinfo{person}{Paul~A Crook}.} \bibinfo{year}{2018}\natexlab{}.
\newblock \showarticletitle{Measuring user satisfaction on smart speaker
  intelligent assistants using intent sensitive query embeddings}. In
  \bibinfo{booktitle}{\emph{Proceedings of the 27th ACM International
  Conference on Information and Knowledge Management}}.
  \bibinfo{pages}{1183--1192}.
\newblock


\bibitem[\protect\citeauthoryear{Kim, Galley, Gunasekara, Lee, Atkinson, Peng,
  Schulz, Gao, Li, Adada, et~al\mbox{.}}{Kim et~al\mbox{.}}{2019}]%
        {kim2019eighth}
\bibfield{author}{\bibinfo{person}{Seokhwan Kim}, \bibinfo{person}{Michel
  Galley}, \bibinfo{person}{Chulaka Gunasekara}, \bibinfo{person}{Sungjin Lee},
  \bibinfo{person}{Adam Atkinson}, \bibinfo{person}{Baolin Peng},
  \bibinfo{person}{Hannes Schulz}, \bibinfo{person}{Jianfeng Gao},
  \bibinfo{person}{Jinchao Li}, \bibinfo{person}{Mahmoud Adada},
  {et~al\mbox{.}}} \bibinfo{year}{2019}\natexlab{}.
\newblock \showarticletitle{The eighth dialog system technology challenge}.
\newblock \bibinfo{journal}{\emph{arXiv preprint arXiv:1911.06394}}
  (\bibinfo{year}{2019}).
\newblock


\bibitem[\protect\citeauthoryear{Kiseleva, Williams, Hassan~Awadallah, Crook,
  Zitouni, and Anastasakos}{Kiseleva et~al\mbox{.}}{2016a}]%
        {kiseleva2016predicting}
\bibfield{author}{\bibinfo{person}{Julia Kiseleva}, \bibinfo{person}{Kyle
  Williams}, \bibinfo{person}{Ahmed Hassan~Awadallah}, \bibinfo{person}{Aidan~C
  Crook}, \bibinfo{person}{Imed Zitouni}, {and} \bibinfo{person}{Tasos
  Anastasakos}.} \bibinfo{year}{2016}\natexlab{a}.
\newblock \showarticletitle{Predicting user satisfaction with intelligent
  assistants}. In \bibinfo{booktitle}{\emph{Proceedings of the 39th
  International ACM SIGIR conference on Research and Development in Information
  Retrieval}}. \bibinfo{pages}{45--54}.
\newblock


\bibitem[\protect\citeauthoryear{Kiseleva, Williams, Jiang, Hassan~Awadallah,
  Crook, Zitouni, and Anastasakos}{Kiseleva et~al\mbox{.}}{2016b}]%
        {kiseleva2016understanding}
\bibfield{author}{\bibinfo{person}{Julia Kiseleva}, \bibinfo{person}{Kyle
  Williams}, \bibinfo{person}{Jiepu Jiang}, \bibinfo{person}{Ahmed
  Hassan~Awadallah}, \bibinfo{person}{Aidan~C Crook}, \bibinfo{person}{Imed
  Zitouni}, {and} \bibinfo{person}{Tasos Anastasakos}.}
  \bibinfo{year}{2016}\natexlab{b}.
\newblock \showarticletitle{Understanding user satisfaction with intelligent
  assistants}. In \bibinfo{booktitle}{\emph{Proceedings of the 2016 ACM on
  Conference on Human Information Interaction and Retrieval}}.
  \bibinfo{pages}{121--130}.
\newblock


\bibitem[\protect\citeauthoryear{Kupiec}{Kupiec}{1992}]%
        {kupiec1992robust}
\bibfield{author}{\bibinfo{person}{Julian Kupiec}.}
  \bibinfo{year}{1992}\natexlab{}.
\newblock \showarticletitle{Robust part-of-speech tagging using a hidden Markov
  model}.
\newblock \bibinfo{journal}{\emph{Computer speech \& language}}
  \bibinfo{volume}{6}, \bibinfo{number}{3} (\bibinfo{year}{1992}),
  \bibinfo{pages}{225--242}.
\newblock


\bibitem[\protect\citeauthoryear{Lan, Mao, Huang, and Wei}{Lan
  et~al\mbox{.}}{2019}]%
        {lan2019talk}
\bibfield{author}{\bibinfo{person}{Tian Lan}, \bibinfo{person}{Xianling Mao},
  \bibinfo{person}{Heyan Huang}, {and} \bibinfo{person}{Wei Wei}.}
  \bibinfo{year}{2019}\natexlab{}.
\newblock \showarticletitle{When to Talk: Chatbot Controls the Timing of
  Talking during Multi-turn Open-domain Dialogue Generation}.
\newblock \bibinfo{journal}{\emph{arXiv preprint arXiv:1912.09879}}
  (\bibinfo{year}{2019}).
\newblock


\bibitem[\protect\citeauthoryear{Lan, Mao, Wei, Gao, and Huang}{Lan
  et~al\mbox{.}}{2020}]%
        {lan2020pone}
\bibfield{author}{\bibinfo{person}{Tian Lan}, \bibinfo{person}{Xian-Ling Mao},
  \bibinfo{person}{Wei Wei}, \bibinfo{person}{Xiaoyan Gao}, {and}
  \bibinfo{person}{Heyan Huang}.} \bibinfo{year}{2020}\natexlab{}.
\newblock \showarticletitle{PONE: A Novel Automatic Evaluation Metric for
  Open-Domain Generative Dialogue Systems}.
\newblock \bibinfo{journal}{\emph{arXiv preprint arXiv:2004.02399}}
  (\bibinfo{year}{2020}).
\newblock


\bibitem[\protect\citeauthoryear{Landauer and Dumais}{Landauer and
  Dumais}{1997}]%
        {landauer1997solution}
\bibfield{author}{\bibinfo{person}{Thomas~K Landauer} {and}
  \bibinfo{person}{Susan~T Dumais}.} \bibinfo{year}{1997}\natexlab{}.
\newblock \showarticletitle{A solution to Plato's problem: The latent semantic
  analysis theory of acquisition, induction, and representation of knowledge.}
\newblock \bibinfo{journal}{\emph{Psychological review}} \bibinfo{volume}{104},
  \bibinfo{number}{2} (\bibinfo{year}{1997}), \bibinfo{pages}{211}.
\newblock


\bibitem[\protect\citeauthoryear{Li, Su, Shen, Li, Cao, and Niu}{Li
  et~al\mbox{.}}{2017}]%
        {li2017dailydialog}
\bibfield{author}{\bibinfo{person}{Yanran Li}, \bibinfo{person}{Hui Su},
  \bibinfo{person}{Xiaoyu Shen}, \bibinfo{person}{Wenjie Li},
  \bibinfo{person}{Ziqiang Cao}, {and} \bibinfo{person}{Shuzi Niu}.}
  \bibinfo{year}{2017}\natexlab{}.
\newblock \showarticletitle{Dailydialog: A manually labelled multi-turn
  dialogue dataset}.
\newblock \bibinfo{journal}{\emph{arXiv preprint arXiv:1710.03957}}
  (\bibinfo{year}{2017}).
\newblock


\bibitem[\protect\citeauthoryear{LIPANI, CARTERETTE, and YILMAZ}{LIPANI
  et~al\mbox{.}}{2021}]%
        {lipani2021doing}
\bibfield{author}{\bibinfo{person}{ALDO LIPANI}, \bibinfo{person}{BEN
  CARTERETTE}, {and} \bibinfo{person}{EMINE YILMAZ}.}
  \bibinfo{year}{2021}\natexlab{}.
\newblock \showarticletitle{How Am I Doing?: Evaluating Conversational Search
  Systems Offline}.
\newblock \bibinfo{journal}{\emph{ACM Transactions on Information Systems}}
  (\bibinfo{year}{2021}).
\newblock


\bibitem[\protect\citeauthoryear{Liu, Lowe, Serban, Noseworthy, Charlin, and
  Pineau}{Liu et~al\mbox{.}}{2016}]%
        {liu2016not}
\bibfield{author}{\bibinfo{person}{Chia-Wei Liu}, \bibinfo{person}{Ryan Lowe},
  \bibinfo{person}{Iulian~V Serban}, \bibinfo{person}{Michael Noseworthy},
  \bibinfo{person}{Laurent Charlin}, {and} \bibinfo{person}{Joelle Pineau}.}
  \bibinfo{year}{2016}\natexlab{}.
\newblock \showarticletitle{How not to evaluate your dialogue system: An
  empirical study of unsupervised evaluation metrics for dialogue response
  generation}.
\newblock \bibinfo{journal}{\emph{arXiv preprint arXiv:1603.08023}}
  (\bibinfo{year}{2016}).
\newblock


\bibitem[\protect\citeauthoryear{Liu, Zhou, and Wilson}{Liu
  et~al\mbox{.}}{2021}]%
        {liu2021meta}
\bibfield{author}{\bibinfo{person}{Zeyang Liu}, \bibinfo{person}{Ke Zhou},
  {and} \bibinfo{person}{Max~L Wilson}.} \bibinfo{year}{2021}\natexlab{}.
\newblock \showarticletitle{Meta-evaluation of Conversational Search Evaluation
  Metrics}.
\newblock \bibinfo{journal}{\emph{arXiv preprint arXiv:2104.13453}}
  (\bibinfo{year}{2021}).
\newblock


\bibitem[\protect\citeauthoryear{Lowe, Noseworthy, Serban, Angelard-Gontier,
  Bengio, and Pineau}{Lowe et~al\mbox{.}}{2017}]%
        {lowe2017towards}
\bibfield{author}{\bibinfo{person}{Ryan Lowe}, \bibinfo{person}{Michael
  Noseworthy}, \bibinfo{person}{Iulian~V Serban}, \bibinfo{person}{Nicolas
  Angelard-Gontier}, \bibinfo{person}{Yoshua Bengio}, {and}
  \bibinfo{person}{Joelle Pineau}.} \bibinfo{year}{2017}\natexlab{}.
\newblock \showarticletitle{Towards an automatic turing test: Learning to
  evaluate dialogue responses}.
\newblock \bibinfo{journal}{\emph{arXiv preprint arXiv:1708.07149}}
  (\bibinfo{year}{2017}).
\newblock


\bibitem[\protect\citeauthoryear{Lowe, Pow, Serban, and Pineau}{Lowe
  et~al\mbox{.}}{2015}]%
        {lowe2015ubuntu}
\bibfield{author}{\bibinfo{person}{Ryan Lowe}, \bibinfo{person}{Nissan Pow},
  \bibinfo{person}{Iulian Serban}, {and} \bibinfo{person}{Joelle Pineau}.}
  \bibinfo{year}{2015}\natexlab{}.
\newblock \showarticletitle{The ubuntu dialogue corpus: A large dataset for
  research in unstructured multi-turn dialogue systems}.
\newblock \bibinfo{journal}{\emph{arXiv preprint arXiv:1506.08909}}
  (\bibinfo{year}{2015}).
\newblock


\bibitem[\protect\citeauthoryear{Mehri and Eskenazi}{Mehri and
  Eskenazi}{2020}]%
        {mehri2020usr}
\bibfield{author}{\bibinfo{person}{Shikib Mehri} {and} \bibinfo{person}{Maxine
  Eskenazi}.} \bibinfo{year}{2020}\natexlab{}.
\newblock \showarticletitle{USR: An Unsupervised and Reference Free Evaluation
  Metric for Dialog Generation}.
\newblock \bibinfo{journal}{\emph{arXiv preprint arXiv:2005.00456}}
  (\bibinfo{year}{2020}).
\newblock


\bibitem[\protect\citeauthoryear{Mitchell and Lapata}{Mitchell and
  Lapata}{2008}]%
        {mitchell2008vector}
\bibfield{author}{\bibinfo{person}{Jeff Mitchell} {and}
  \bibinfo{person}{Mirella Lapata}.} \bibinfo{year}{2008}\natexlab{}.
\newblock \showarticletitle{Vector-based models of semantic composition}. In
  \bibinfo{booktitle}{\emph{proceedings of ACL-08: HLT}}.
  \bibinfo{pages}{236--244}.
\newblock


\bibitem[\protect\citeauthoryear{Novikova, Du{\v{s}}ek, Curry, and
  Rieser}{Novikova et~al\mbox{.}}{2017}]%
        {novikova2017we}
\bibfield{author}{\bibinfo{person}{Jekaterina Novikova},
  \bibinfo{person}{Ond{\v{r}}ej Du{\v{s}}ek}, \bibinfo{person}{Amanda~Cercas
  Curry}, {and} \bibinfo{person}{Verena Rieser}.}
  \bibinfo{year}{2017}\natexlab{}.
\newblock \showarticletitle{Why we need new evaluation metrics for NLG}.
\newblock \bibinfo{journal}{\emph{arXiv preprint arXiv:1707.06875}}
  (\bibinfo{year}{2017}).
\newblock


\bibitem[\protect\citeauthoryear{Papineni, Roukos, Ward, and Zhu}{Papineni
  et~al\mbox{.}}{2002}]%
        {papineni2002bleu}
\bibfield{author}{\bibinfo{person}{Kishore Papineni}, \bibinfo{person}{Salim
  Roukos}, \bibinfo{person}{Todd Ward}, {and} \bibinfo{person}{Wei-Jing Zhu}.}
  \bibinfo{year}{2002}\natexlab{}.
\newblock \showarticletitle{BLEU: a method for automatic evaluation of machine
  translation}. In \bibinfo{booktitle}{\emph{Proceedings of the 40th annual
  meeting on association for computational linguistics}}. Association for
  Computational Linguistics, \bibinfo{pages}{311--318}.
\newblock


\bibitem[\protect\citeauthoryear{Qu, Yang, Croft, Trippas, Zhang, and Qiu}{Qu
  et~al\mbox{.}}{2018}]%
        {qu2018analyzing}
\bibfield{author}{\bibinfo{person}{Chen Qu}, \bibinfo{person}{Liu Yang},
  \bibinfo{person}{W~Bruce Croft}, \bibinfo{person}{Johanne~R Trippas},
  \bibinfo{person}{Yongfeng Zhang}, {and} \bibinfo{person}{Minghui Qiu}.}
  \bibinfo{year}{2018}\natexlab{}.
\newblock \showarticletitle{Analyzing and characterizing user intent in
  information-seeking conversations}. In \bibinfo{booktitle}{\emph{The 41st
  International ACM SIGIR Conference on Research \& Development in Information
  Retrieval}}. ACM, \bibinfo{pages}{989--992}.
\newblock


\bibitem[\protect\citeauthoryear{Radlinski and Craswell}{Radlinski and
  Craswell}{2017}]%
        {radlinski2017theoretical}
\bibfield{author}{\bibinfo{person}{Filip Radlinski} {and} \bibinfo{person}{Nick
  Craswell}.} \bibinfo{year}{2017}\natexlab{}.
\newblock \showarticletitle{A theoretical framework for conversational search}.
  In \bibinfo{booktitle}{\emph{Proceedings of the 2017 conference on conference
  human information interaction and retrieval}}. ACM,
  \bibinfo{pages}{117--126}.
\newblock


\bibitem[\protect\citeauthoryear{Rajpurkar, Zhang, Lopyrev, and
  Liang}{Rajpurkar et~al\mbox{.}}{2016}]%
        {rajpurkar2016squad}
\bibfield{author}{\bibinfo{person}{Pranav Rajpurkar}, \bibinfo{person}{Jian
  Zhang}, \bibinfo{person}{Konstantin Lopyrev}, {and} \bibinfo{person}{Percy
  Liang}.} \bibinfo{year}{2016}\natexlab{}.
\newblock \showarticletitle{Squad: 100,000+ questions for machine comprehension
  of text}.
\newblock \bibinfo{journal}{\emph{arXiv preprint arXiv:1606.05250}}
  (\bibinfo{year}{2016}).
\newblock


\bibitem[\protect\citeauthoryear{Ratnaparkhi}{Ratnaparkhi}{1996}]%
        {ratnaparkhi1996maximum}
\bibfield{author}{\bibinfo{person}{Adwait Ratnaparkhi}.}
  \bibinfo{year}{1996}\natexlab{}.
\newblock \showarticletitle{A maximum entropy model for part-of-speech
  tagging}. In \bibinfo{booktitle}{\emph{Conference on empirical methods in
  natural language processing}}.
\newblock


\bibitem[\protect\citeauthoryear{Rus and Lintean}{Rus and Lintean}{2012}]%
        {rus2012comparison}
\bibfield{author}{\bibinfo{person}{Vasile Rus} {and} \bibinfo{person}{Mihai
  Lintean}.} \bibinfo{year}{2012}\natexlab{}.
\newblock \showarticletitle{A comparison of greedy and optimal assessment of
  natural language student input using word-to-word similarity metrics}. In
  \bibinfo{booktitle}{\emph{Proceedings of the Seventh Workshop on Building
  Educational Applications Using NLP}}. Association for Computational
  Linguistics, \bibinfo{pages}{157--162}.
\newblock


\bibitem[\protect\citeauthoryear{Sakai}{Sakai}{2006}]%
        {sakai2006evaluating}
\bibfield{author}{\bibinfo{person}{Tetsuya Sakai}.}
  \bibinfo{year}{2006}\natexlab{}.
\newblock \showarticletitle{Evaluating evaluation metrics based on the
  bootstrap}. In \bibinfo{booktitle}{\emph{Proceedings of the 29th annual
  international ACM SIGIR conference on Research and development in information
  retrieval}}. ACM, \bibinfo{pages}{525--532}.
\newblock


\bibitem[\protect\citeauthoryear{Sakai}{Sakai}{2007}]%
        {sakai2007reliability}
\bibfield{author}{\bibinfo{person}{Tetsuya Sakai}.}
  \bibinfo{year}{2007}\natexlab{}.
\newblock \showarticletitle{On the reliability of information retrieval metrics
  based on graded relevance}.
\newblock \bibinfo{journal}{\emph{Information processing \& management}}
  \bibinfo{volume}{43}, \bibinfo{number}{2} (\bibinfo{year}{2007}),
  \bibinfo{pages}{531--548}.
\newblock


\bibitem[\protect\citeauthoryear{Sakai}{Sakai}{2012}]%
        {sakai2012evaluation}
\bibfield{author}{\bibinfo{person}{Tetsuya Sakai}.}
  \bibinfo{year}{2012}\natexlab{}.
\newblock \showarticletitle{Evaluation with informational and navigational
  intents}. In \bibinfo{booktitle}{\emph{Proceedings of the 21st international
  conference on World Wide Web}}. ACM, \bibinfo{pages}{499--508}.
\newblock


\bibitem[\protect\citeauthoryear{Sakai}{Sakai}{2013}]%
        {sakai2013metrics}
\bibfield{author}{\bibinfo{person}{Tetsuya Sakai}.}
  \bibinfo{year}{2013}\natexlab{}.
\newblock \showarticletitle{Metrics, statistics, tests}. In
  \bibinfo{booktitle}{\emph{PROMISE winter school}}. Springer,
  \bibinfo{pages}{116--163}.
\newblock


\bibitem[\protect\citeauthoryear{Sakai et~al\mbox{.}}{Sakai
  et~al\mbox{.}}{2005}]%
        {sakai2005effect}
\bibfield{author}{\bibinfo{person}{Tetsuya Sakai} {et~al\mbox{.}}}
  \bibinfo{year}{2005}\natexlab{}.
\newblock \showarticletitle{The Effect of Topic Sampling on Sensitivity
  Comparisons of Information Retrieval Metrics.}. In
  \bibinfo{booktitle}{\emph{Proceedings of NTCIR-5}}.
\newblock


\bibitem[\protect\citeauthoryear{Sakai and Kando}{Sakai and Kando}{2008}]%
        {sakai2008information}
\bibfield{author}{\bibinfo{person}{Tetsuya Sakai} {and} \bibinfo{person}{Noriko
  Kando}.} \bibinfo{year}{2008}\natexlab{}.
\newblock \showarticletitle{On information retrieval metrics designed for
  evaluation with incomplete relevance assessments}.
\newblock \bibinfo{journal}{\emph{Information Retrieval}} \bibinfo{volume}{11},
  \bibinfo{number}{5} (\bibinfo{year}{2008}), \bibinfo{pages}{447--470}.
\newblock


\bibitem[\protect\citeauthoryear{Sakai and Zeng}{Sakai and Zeng}{2019}]%
        {sakai2019diversity}
\bibfield{author}{\bibinfo{person}{Tetsuya Sakai} {and}
  \bibinfo{person}{Zhaohao Zeng}.} \bibinfo{year}{2019}\natexlab{}.
\newblock \showarticletitle{Which Diversity Evaluation Measures Are" Good"?}.
  In \bibinfo{booktitle}{\emph{Proceedings of the 42nd International ACM SIGIR
  Conference on Research and Development in Information Retrieval}}.
  \bibinfo{pages}{595--604}.
\newblock


\bibitem[\protect\citeauthoryear{Sanderson, Paramita, Clough, and
  Kanoulas}{Sanderson et~al\mbox{.}}{2010}]%
        {sanderson2010user}
\bibfield{author}{\bibinfo{person}{Mark Sanderson},
  \bibinfo{person}{Monica~Lestari Paramita}, \bibinfo{person}{Paul Clough},
  {and} \bibinfo{person}{Evangelos Kanoulas}.} \bibinfo{year}{2010}\natexlab{}.
\newblock \showarticletitle{Do user preferences and evaluation measures line
  up?}. In \bibinfo{booktitle}{\emph{Proceedings of the 33rd international ACM
  SIGIR conference on Research and development in information retrieval}}. ACM,
  \bibinfo{pages}{555--562}.
\newblock


\bibitem[\protect\citeauthoryear{Serban, Sordoni, Lowe, Charlin, Pineau,
  Courville, and Bengio}{Serban et~al\mbox{.}}{2017}]%
        {serban2017hierarchical}
\bibfield{author}{\bibinfo{person}{Iulian~Vlad Serban},
  \bibinfo{person}{Alessandro Sordoni}, \bibinfo{person}{Ryan Lowe},
  \bibinfo{person}{Laurent Charlin}, \bibinfo{person}{Joelle Pineau},
  \bibinfo{person}{Aaron Courville}, {and} \bibinfo{person}{Yoshua Bengio}.}
  \bibinfo{year}{2017}\natexlab{}.
\newblock \showarticletitle{A hierarchical latent variable encoder-decoder
  model for generating dialogues}. In \bibinfo{booktitle}{\emph{Thirty-First
  AAAI Conference on Artificial Intelligence}}.
\newblock


\bibitem[\protect\citeauthoryear{Sidorov, Gelbukh, G{\'o}mez-Adorno, and
  Pinto}{Sidorov et~al\mbox{.}}{2014}]%
        {sidorov2014soft}
\bibfield{author}{\bibinfo{person}{Grigori Sidorov}, \bibinfo{person}{Alexander
  Gelbukh}, \bibinfo{person}{Helena G{\'o}mez-Adorno}, {and}
  \bibinfo{person}{David Pinto}.} \bibinfo{year}{2014}\natexlab{}.
\newblock \showarticletitle{Soft similarity and soft cosine measure: Similarity
  of features in vector space model}.
\newblock \bibinfo{journal}{\emph{Computaci{\'o}n y Sistemas}}
  \bibinfo{volume}{18}, \bibinfo{number}{3} (\bibinfo{year}{2014}),
  \bibinfo{pages}{491--504}.
\newblock


\bibitem[\protect\citeauthoryear{Sinha, Parthasarathi, Wang, Lowe, Hamilton,
  and Pineau}{Sinha et~al\mbox{.}}{2020}]%
        {sinha2020learning}
\bibfield{author}{\bibinfo{person}{Koustuv Sinha}, \bibinfo{person}{Prasanna
  Parthasarathi}, \bibinfo{person}{Jasmine Wang}, \bibinfo{person}{Ryan Lowe},
  \bibinfo{person}{William~L Hamilton}, {and} \bibinfo{person}{Joelle Pineau}.}
  \bibinfo{year}{2020}\natexlab{}.
\newblock \showarticletitle{Learning an Unreferenced Metric for Online Dialogue
  Evaluation}.
\newblock \bibinfo{journal}{\emph{arXiv preprint arXiv:2005.00583}}
  (\bibinfo{year}{2020}).
\newblock


\bibitem[\protect\citeauthoryear{Smucker, Allan, and Carterette}{Smucker
  et~al\mbox{.}}{2007}]%
        {smucker2007comparison}
\bibfield{author}{\bibinfo{person}{Mark~D Smucker}, \bibinfo{person}{James
  Allan}, {and} \bibinfo{person}{Ben Carterette}.}
  \bibinfo{year}{2007}\natexlab{}.
\newblock \showarticletitle{A comparison of statistical significance tests for
  information retrieval evaluation}. In \bibinfo{booktitle}{\emph{Proceedings
  of the sixteenth ACM conference on Conference on information and knowledge
  management}}. \bibinfo{pages}{623--632}.
\newblock


\bibitem[\protect\citeauthoryear{Tao, Mou, Zhao, and Yan}{Tao
  et~al\mbox{.}}{2018}]%
        {tao2018ruber}
\bibfield{author}{\bibinfo{person}{Chongyang Tao}, \bibinfo{person}{Lili Mou},
  \bibinfo{person}{Dongyan Zhao}, {and} \bibinfo{person}{Rui Yan}.}
  \bibinfo{year}{2018}\natexlab{}.
\newblock \showarticletitle{Ruber: An unsupervised method for automatic
  evaluation of open-domain dialog systems}. In
  \bibinfo{booktitle}{\emph{Thirty-Second AAAI Conference on Artificial
  Intelligence}}.
\newblock


\bibitem[\protect\citeauthoryear{Vtyurina, Savenkov, Agichtein, and
  Clarke}{Vtyurina et~al\mbox{.}}{2017}]%
        {vtyurina2017exploring}
\bibfield{author}{\bibinfo{person}{Alexandra Vtyurina}, \bibinfo{person}{Denis
  Savenkov}, \bibinfo{person}{Eugene Agichtein}, {and}
  \bibinfo{person}{Charles~LA Clarke}.} \bibinfo{year}{2017}\natexlab{}.
\newblock \showarticletitle{Exploring conversational search with humans,
  assistants, and wizards}. In \bibinfo{booktitle}{\emph{Proceedings of the
  2017 chi conference extended abstracts on human factors in computing
  systems}}. \bibinfo{pages}{2187--2193}.
\newblock


\bibitem[\protect\citeauthoryear{Yang, Yih, and Meek}{Yang
  et~al\mbox{.}}{2015}]%
        {yang2015wikiqa}
\bibfield{author}{\bibinfo{person}{Yi Yang}, \bibinfo{person}{Wen-tau Yih},
  {and} \bibinfo{person}{Christopher Meek}.} \bibinfo{year}{2015}\natexlab{}.
\newblock \showarticletitle{Wikiqa: A challenge dataset for open-domain
  question answering}. In \bibinfo{booktitle}{\emph{Proceedings of the 2015
  conference on empirical methods in natural language processing}}.
  \bibinfo{pages}{2013--2018}.
\newblock


\bibitem[\protect\citeauthoryear{Ye, Manotumruksa, and Yilmaz}{Ye
  et~al\mbox{.}}{2021}]%
        {ye2021multiwoz}
\bibfield{author}{\bibinfo{person}{Fanghua Ye}, \bibinfo{person}{Jarana
  Manotumruksa}, {and} \bibinfo{person}{Emine Yilmaz}.}
  \bibinfo{year}{2021}\natexlab{}.
\newblock \showarticletitle{MultiWOZ 2.4: A Multi-Domain Task-Oriented Dialogue
  Dataset with Essential Annotation Corrections to Improve State Tracking
  Evaluation}.
\newblock \bibinfo{journal}{\emph{arXiv preprint arXiv:2104.00773}}
  (\bibinfo{year}{2021}).
\newblock


\bibitem[\protect\citeauthoryear{Yuma, Yoshinaga, and Toyoda}{Yuma
  et~al\mbox{.}}{2020}]%
        {yuma2020ubleu}
\bibfield{author}{\bibinfo{person}{Tsuta Yuma}, \bibinfo{person}{Naoki
  Yoshinaga}, {and} \bibinfo{person}{Masashi Toyoda}.}
  \bibinfo{year}{2020}\natexlab{}.
\newblock \showarticletitle{uBLEU: Uncertainty-Aware Automatic Evaluation
  Method for Open-Domain Dialogue Systems}. In
  \bibinfo{booktitle}{\emph{Proceedings of the 58th Annual Meeting of the
  Association for Computational Linguistics: Student Research Workshop}}.
  \bibinfo{pages}{199--206}.
\newblock


\bibitem[\protect\citeauthoryear{Zhang, Dinan, Urbanek, Szlam, Kiela, and
  Weston}{Zhang et~al\mbox{.}}{2018}]%
        {zhang2018personalizing}
\bibfield{author}{\bibinfo{person}{Saizheng Zhang}, \bibinfo{person}{Emily
  Dinan}, \bibinfo{person}{Jack Urbanek}, \bibinfo{person}{Arthur Szlam},
  \bibinfo{person}{Douwe Kiela}, {and} \bibinfo{person}{Jason Weston}.}
  \bibinfo{year}{2018}\natexlab{}.
\newblock \showarticletitle{Personalizing dialogue agents: I have a dog, do you
  have pets too?}
\newblock \bibinfo{journal}{\emph{arXiv preprint arXiv:1801.07243}}
  (\bibinfo{year}{2018}).
\newblock


\bibitem[\protect\citeauthoryear{Zhang*, Kishore*, Wu*, Weinberger, and
  Artzi}{Zhang* et~al\mbox{.}}{2020}]%
        {bert-score}
\bibfield{author}{\bibinfo{person}{Tianyi Zhang*}, \bibinfo{person}{Varsha
  Kishore*}, \bibinfo{person}{Felix Wu*}, \bibinfo{person}{Kilian~Q.
  Weinberger}, {and} \bibinfo{person}{Yoav Artzi}.}
  \bibinfo{year}{2020}\natexlab{}.
\newblock \showarticletitle{BERTScore: Evaluating Text Generation with BERT}.
  In \bibinfo{booktitle}{\emph{International Conference on Learning
  Representations}}.
\newblock
\urldef\tempurl%
\url{https://openreview.net/forum?id=SkeHuCVFDr}
\showURL{%
\tempurl}


\bibitem[\protect\citeauthoryear{Zhou, Cummins, Lalmas, and Jose}{Zhou
  et~al\mbox{.}}{2012}]%
        {zhou2012evaluating}
\bibfield{author}{\bibinfo{person}{Ke Zhou}, \bibinfo{person}{Ronan Cummins},
  \bibinfo{person}{Mounia Lalmas}, {and} \bibinfo{person}{Joemon~M Jose}.}
  \bibinfo{year}{2012}\natexlab{}.
\newblock \showarticletitle{Evaluating aggregated search pages}. In
  \bibinfo{booktitle}{\emph{Proceedings of the 35th international ACM SIGIR
  conference on Research and development in information retrieval}}. ACM,
  \bibinfo{pages}{115--124}.
\newblock


\end{thebibliography}
